\documentclass[twocolumn,showpacs,preprintnumbers,amsmath,amssymb,pra,aps,superscriptaddress]{revtex4-1}

\usepackage{graphicx}
\usepackage{amsmath}
\usepackage{verbatim}
\usepackage{bm}
\usepackage{siunitx}
\usepackage[space]{grffile}
\usepackage{xr} 
\usepackage{color}

\newcommand{\Dzero}{D_0}
\newcommand{\Done}{D_1}

\newcommand{\Qone}{Q_1}
\newcommand{\Qtwo}{Q_2}
\newcommand{\Dall}{D_{\mathrm{U}}}
\newcommand{\taucycle}{\tau_{\mathrm{cycle}}}
\newcommand{\types}{{s}}
\newcommand{\typed}{{d}}

\newcommand{\ns}{\mathrm{ns}}
\newcommand{\us}{\mu\mathrm{s}}
\newcommand{\GHz}{\mathrm{GHz}}
\newcommand{\MHz}{\mathrm{MHz}}
\newcommand{\MSps}{\mathrm{MSamples/s}}
\newcommand{\dB}{\mathrm{dB}}
\newcommand{\dBm}{\mathrm{dBm}}
\newcommand{\ket}[1]{\left\lvert #1 \right\rangle}
\newcommand{\Fa}{\mathcal{F}_{\mathrm{a}}}
\newcommand{\Fd}{\mathcal{F}_{\mathrm{d}}}
\newcommand{\Tone}{T_{1}}

\newcommand{\Techo}{T_2^\mathrm{echo}}

\newcommand{\Prf}{P_{\mathrm{rf}}}
\newcommand{\frf}{f_{\mathrm{rf}}}
\newcommand{\tauint}{\tau_\mathrm{int}}
\newcommand{\taumeas}{\tau_\mathrm{r}}
\newcommand{\taud}{\tau_\mathrm{d}}
\newcommand{\taup}{\tau_\mathrm{p}}

\newcommand{\RTE}{\mathrm{RTE}}
\newcommand{\RTEmean}{\overline{\mathrm{RTE}}}
\newcommand{\Fidone}{F_{1}}
\newcommand{\AllXYerror}{\mathcal{E}_{\mathrm{AllXY}}}
\newcommand{\nphoton}{\overline{n}}
\newcommand{\ncrit}{n_\mathrm{crit}}
\newcommand{\res}{\mathrm{r}}
\newcommand{\frzero}{f_{\res,\ket{0}}}
\newcommand{\frone}{f_{\res,\ket{1}}}
\newcommand{\fbare}{f_{\res, \mathrm{bare}}}
\newcommand{\fqubit}{f_{\mathrm{q}}}
\newcommand{\psingle}{p_{\mathrm{s}}}
\newcommand{\Dd}{\mathcal{D}}
\newcommand{\rhoqb}{\rho^{\mathrm{qb}}}

\begin{document}

\title{Active resonator reset in the nonlinear dispersive regime of circuit QED}
\author{C.~C.~Bultink}
\author{M.~A.~Rol}
\affiliation{QuTech, Delft University of Technology, P.O. Box 5046, 2600 GA Delft, The Netherlands}
\affiliation{Kavli Institute of Nanoscience, Delft University of Technology, P.O. Box 5046, 2600 GA Delft, The Netherlands}
\author{T.~E.~O'Brien}
\affiliation{Instituut-Lorentz for Theoretical Physics, Leiden University, Leiden, The Netherlands}
\author{X.~Fu}
\affiliation{QuTech, Delft University of Technology, P.O. Box 5046, 2600 GA Delft, The Netherlands}
\affiliation{College of Computer, National University of Defense Technology, Changsha, China 410073}
\author{B.~C.~S.~Dikken}
\author{C.~Dickel}
\author{R.~F.~L.~Vermeulen}
\affiliation{QuTech, Delft University of Technology, P.O. Box 5046, 2600 GA Delft, The Netherlands}
\affiliation{Kavli Institute of Nanoscience, Delft University of Technology, P.O. Box 5046, 2600 GA Delft, The Netherlands}

\author{J.~C.~de~Sterke}
\affiliation{Topic Embedded Systems B.V., P.O. Box 440, 5680 AK Best, The Netherlands}
\affiliation{QuTech, Delft University of Technology, P.O. Box 5046, 2600 GA Delft, The Netherlands}
\author{A.~Bruno}

\author{R.~N.~Schouten}
\author{L.~DiCarlo}
\affiliation{QuTech, Delft University of Technology, P.O. Box 5046, 2600 GA Delft, The Netherlands}
\affiliation{Kavli Institute of Nanoscience, Delft University of Technology, P.O. Box 5046, 2600 GA Delft, The Netherlands}

\date{\today}

\begin{abstract}
We present two pulse schemes for actively depleting measurement photons from a readout resonator in the nonlinear dispersive regime of circuit QED.
One method uses digital feedback conditioned on the measurement outcome while the other is unconditional.
In the absence of analytic forms and symmetries to exploit in this nonlinear regime, the depletion pulses are numerically optimized using the Powell method.
We shorten the photon depletion time by more than six inverse resonator linewidths compared to passive depletion by waiting. 
We quantify the benefit by emulating an ancilla qubit performing repeated quantum parity checks in a repetition code. Fast depletion increases the mean number of cycles to a spurious error detection event from order 1 to 75 at a $1~\us$ cycle time.

\end{abstract}

\maketitle

\begin{figure}[!htb]
  \centering
  \includegraphics[width=\columnwidth]{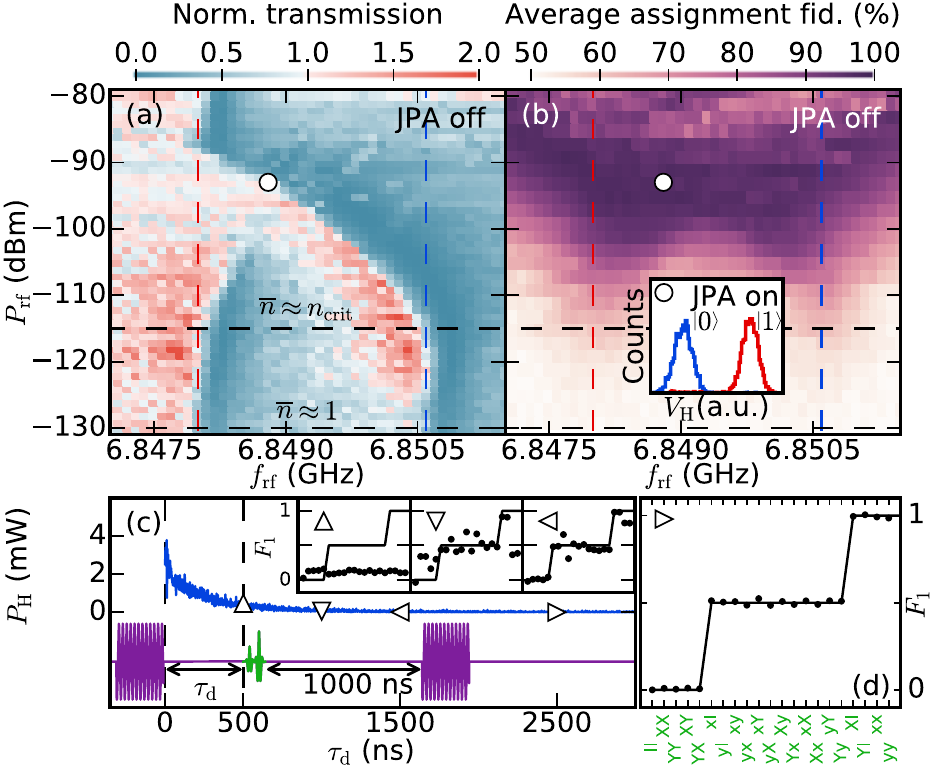}
  \caption{(Color online)
  {\bf Dispersive qubit readout in the nonlinear regime and qubit errors produced by leftover measurement photons.}
  (a) CW feedline transmission spectroscopy as a function of incident power and frequency near the low- and high-power fundamentals of the readout resonator.
  The qubit is simultaneously driven with a weakly saturating CW tone.
  The right (left) vertical line indicates the fundamental $\frzero$ ($\frone$) for qubit in $\ket{0}$ ($\ket{1}$) in the linear regime.
  The dot [also in (b)] indicates the settings $(\Prf,\frf)=(-93~\dBm, 6.8488~\GHz)$ used throughout the experiment.
  (b) Map of average assignment fidelity $\Fa$ as a function of $\Prf$ and $\frf$ with the JPA off ($\taumeas=1200~\ns$, $\tauint=1500~\ns$), obtained from histograms with 4000 shots per qubit state. Inset: Turning on the JPA achieves $\Fa=98.8\%$ with $\taumeas=300~\ns$ and an optimized weight function integrating $\tauint=400~\ns$.
  (c) Illustration of qubits errors induced by leftover photons.
  At $\taud$, after an initial measurement pulse ends, AllXY qubit pulse pairs are applied and a final measurement is performed $1000~\ns$ later to measure $\Fidone$.
  The transient of the decaying homodyne signal, $P_\mathrm{H}$, fits a single-photon relaxation time $1/\kappa=250~\ns$.
  Insets and (d): $\Fidone$ versus pulse pair for several $\taud$. The ideal two-step signature is observed only at $\taud \gtrsim 2500~\ns$.}
  \label{fig:fig1}
\end{figure}
Many protocols in quantum information processing require interleaving qubit gates and measurements in rapid succession.
For example, current experimental implementations of quantum error correction (QEC) schemes~\cite{Reed12,Kelly15,Riste15,Cramer15,Nigg14,Corcoles15,Ofek16} rely on repeated measurements of ancilla qubits to discretize and track errors in the data-carrying part of the system. 
Minimizing the QEC cycle time is essential to avoid build-up of errors beyond the threshold for fault-tolerance.

An attractive architecture for QEC codes is circuit quantum electrodynamics (cQED)~\cite{Blais04}. 
Initially implemented with superconducting qubits, this scheme has since grown to include both semiconducting~\cite{Petersson12} and hybrid qubit platforms~\cite{Larsen15, deLange15}. 
Readout in cQED involves dispersively coupling the qubit to a microwave-frequency resonator causing a qubit-state dependent shift of the fundamental resonance.
This shift can be measured by injecting the resonator with a microwave photon pulse. 
Inversely however, there is a dual sensitivity of the qubit transition frequency to resonator photons (AC Stark shift~\cite{Blais04}), leading to qubit dephasing and detuning, as well as gate errors.
To ensure photons leave the resonator before gates recommence, cQED implementations of QEC include a waiting step after measurement. 
During this dead time, which lasts a significant fraction of the QEC cycle, qubits are susceptible to decoherence. 
Whilst many prerequisites of measurement in cQED devices for QEC have already been demonstrated (including frequency-multiplexed readout via a common feedline~\cite{Groen13}, the use of parametric amplifiers to improve speed and readout fidelity~\cite{Johnson12,Riste12} and null back-action on untargeted qubits~\cite{Saira14}),
comparatively little attention has been given to the fast depletion of resonator photons post measurement.

Two compatible approaches to accelerate photon depletion have been explored.
The first increases the resonator linewidth $\kappa$ while adding a Purcell filter~\cite{Reed10b,Jeffrey14,Kelly15} to avoid enhanced qubit relaxation via the Purcell effect~\cite{Houck08}.
However, increasing $\kappa$ enhances the rate of qubit dephasing due to stray photons~\cite{Sears12, Jin15}, introducing a compromise.
The second approach is to actively deplete photons using a counter pulse, as recently demonstrated by McClure \textit{et al.}~\cite{McClure16}.
This demonstration exploited useful symmetries available when the resonator response is linear. However, reaching the single-shot readout fidelity required for QEC often involves driving the resonator deep into the nonlinear regime, where no such symmetries are available.

In this Letter, we propose and demonstrate two methods of active photon depletion in the nonlinear dispersive regime of cQED. 
The first uses a homebuilt feedback controller to send one of two depletion pulses conditioned on the declared measurement outcome. The second applies a universal pulse independent of measurement outcome. 
We maximize readout fidelity at a measurement power two orders of magnitude larger than that inducing the critical photon number in the resonator~\cite{Blais04}. 
Without analytic expressions and convenient symmetries for this nonlinear regime, we rely exclusively on numerical optimization by Powell's method~\cite{Powell64} to tune up pulses with physically-motivated shapes, defined by two or four parameters.
Both methods shorten the photon depletion by at least $5/\kappa$ compared to depletion by waiting.
We illustrate the benefits of active photon depletion using an emulation of multi-round quantum error correction.
Specifically, we emulate an ancilla qubit performing parity checks~\cite{Saira14,Chow14} by subjecting our qubit to repeated rounds of coherent operations and measurement.
We quantify performance by extracting the mean number of rounds to a measurement outcome that deviates from the ideal result (i.e., an error detection event).
With active depletion, we observe an increase in this mean rounds to event, $\RTEmean$, from $15$ to $39$ due to the reduction of total cycle time to $1~\us \sim 4/\kappa$. 
By further fixing the ancilla to remain in the ground state, $\RTEmean$ increases to $75$. Numerical simulations~\cite{oBrien16} indicate that, when including the same intrinsic coherence for surrounding data qubits, a 5-qubit repetition code (studied in~\cite{Kelly15}) would have a logical error rate below its pseudo-threshold~\cite{Tomita14}.

We employ a 2D cQED chip containing ten transmon qubits with dedicated readout resonators, all capacitively coupled to a common feedline through which all microwave control and measurement pulses are applied.
We focus on one qubit-resonator pair for all data presented. This qubit is operated at its flux sweetspot, with transition from ground ($\ket{0}$) to first-excited ($\ket{1}$) state at $\fqubit=6.477~\GHz$, and average relaxation and Hahn echo times $\Tone=25~\us$ and $\Techo=39~\us$.
The dispersively coupled resonator has a low-power fundamental at $\frzero=6.8506~\GHz$ ($\frone=6.8480~\GHz$) for qubit in $\ket{0}$ ($\ket{1}$), making the dispersive shift $\chi/\pi=-2.6~\MHz$. 
Note that this shift also corresponds to the qubit detuning induced per resonator photon. 
The fundamentals converge to the bare resonator frequency, $\fbare=6.8478~\GHz$, at incident power $\Prf \gtrsim -88~\dBm$. We calibrate a single-photon power $\Prf=-130~\dBm$ using photon-number splitting experiments~\cite{Schuster07,SOMdepletion} and a critical photon number~\cite{Blais04} $\ncrit=(\Delta^2/4g^2)\approx 33$ ($\Prf \approx -115~\dBm$) using $\frzero-\fbare=g^2/2\pi\Delta$ and $\Delta=2\pi(\fqubit-\fbare)$.

Our first objective is to maximize the average assignment fidelity of single-shot readout,
\[
\Fa=1-\frac{1}{2}\left(\epsilon_{01} + \epsilon_{10}\right),
\]
where $\epsilon_{ij}$ is the probability of incorrectly assigning measurement result $j$ for input state $\ket{i}$, $i,j\in\{0,1\}$.
We map $\Fa$ as a function of the power $\Prf$ and frequency $\frf$ of a measurement pulse of duration $\taumeas=1200~\ns$ [Fig.~\ref{fig:fig1}(b)].
$\Fa$ is maximized at an intermediate $\Prf=-93~\dBm$, $22~\dB$ stronger than the $\ncrit$ power.
The nonlinearity is evidenced by the bending of resonator lineshapes in the accompanying continuous-wave (CW) transmission spectroscopy [Fig.~\ref{fig:fig1}(a)].  
We make two additions to further improve $\Fa$. First, we turn on a Josephson parametric amplifier (JPA) as the front-end of our amplification chain, operating in non-degenerate mode with $14~\dB$ of gain.
The improved signal-to-noise ratio allows shortening $\taumeas$ to $300~\ns$. 
Second, we use an optimized weight function (duration $\tauint=400~\ns$) to integrate the demodulated homodyne signal before thresholding.
This weight function consists of the difference of the averaged transients for $\ket{0}$ and for $\ket{1}$~\cite{Ryan15, Magesan15}.
These additions achieve $\Fa=98.8\%$, with $\epsilon_{01}=0.1\%$ and $\epsilon_{10}=2.3\%$ [Inset, Fig.~\ref{fig:fig1}(b)], limited by $\Tone$.

The effect of photons leftover from this strong measurement is conveniently illustrated with a modified AllXY sequence~\cite{Chow10b, ReedPhD13}.
AllXY consists of 21 sequences, each comprised of one pair of pulses [Fig.~\ref{fig:fig1}(d)] applied to the qubit followed by qubit measurement.
The qubit pulses are drawn from the set $\{I,X,Y,x,y\}$, where $I$ denotes the identity, and $X$ and $Y$ $(x$ and $y)$ denote $\pi$ ($\pi/2$) pulses around the $x$ and $y$ axis of the Bloch sphere, respectively.
Ideal pulses leave the qubit in $\ket{0}$ (first 5 pairs), on the equator of the Bloch sphere (next 12), and in $\ket{1}$ (final 4), producing a characteristic two-step signature in the fidelity to $\ket{1}$, $\Fidone$ [Fig.~\ref{fig:fig1}(d)].
The chosen order of pulse pairs reveals clear signatures of errors in many gate parameters~\cite{ReedPhD13}.
Here, we modify the AllXY sequence by applying a measurement pulse ending at a time $\taud$ before the start of the qubit pulses.
The effect of leftover photons on the pulses is clearly visible in Fig.~\ref{fig:fig1}(c). At $\taud \gtrsim 10/\kappa$, $\Fidone$ displays the expected double step.
At $\taud\sim 7/\kappa$, the characteristic signature of moderate qubit detuning is observed in the high/low response of pulse pairs $x$-$y$ and $y$-$x$.
At $\taud\leq 2/\kappa$, the detuning is significant with respect to the Rabi frequency of pulses, which thus barely excite the qubit.

\begin{figure}
  \centering
  \includegraphics[width=\columnwidth]{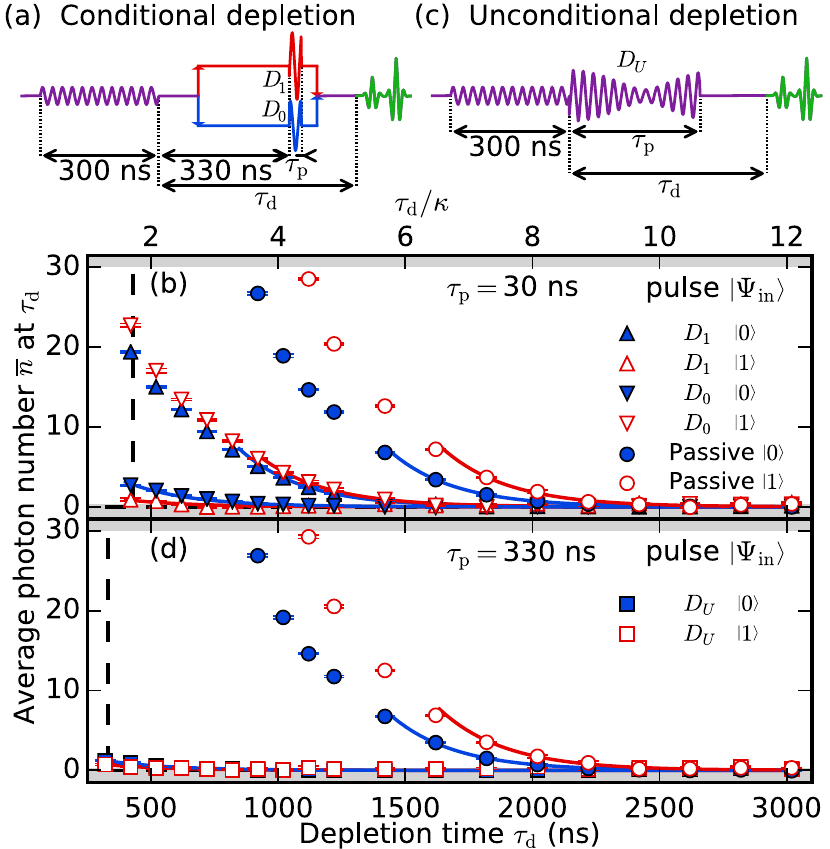}
  \caption{(Color online) {\bf Two active methods of photon depletion compared to passive depletion.}
  (a) Pulse scheme for conditional photon depletion. The controller applies a depletion pulse $\Dzero$ (at $\frzero$) or $\Done$ (at $\frone$), each with separate amplitude and phase, depending on its declared measurement outcome.
  (b) Performance of conditional depletion. Average photon number $\nphoton$ as a function of $\taud$ for all combinations of input qubit state and depletion pulse. Compared to waiting, conditional depletion saves $\geq 1240~(1790)~\ns$ for correct declaration 0 (1).
  (c) Pulse scheme for unconditional active depletion.
  The single depletion pulse $\Dall$, immediately following the nominal measurement pulse, has four parameters corresponding to the amplitude and phase of pulse components at $\frzero$ and $\frone$.
  (d) Performance of unconditional depletion. Unconditional depletion saves $\geq1650~(1920)~\ns$ for $\ket{0}$ ($\ket{1}$). Exponential best fits (curves) to the data in the linear regime  ($\nphoton\leq8$) give
  $1/\kappa=255\pm5~\ns$.}
  \label{fig:fig2}
\end{figure}

We now focus on the calibration of AllXY as a photon detector, suitable for the optimization of depletion pulses. 
Because we miss analytic formulas in the nonlinear regime, pulse optimization relies on numerical minimization of the residual average photon number $\nphoton$ using Powell's method.
We choose $\AllXYerror$ as cost function, defined as the sum of the absolute deviations from the ideal-result fit.
We find experimentally that $\AllXYerror=\alpha\nphoton(\taud)+\beta$ for $\nphoton \lesssim 30$.
The calibration of coefficients $\alpha$ and $\beta$ is described in~\cite{SOMdepletion}.
Measurement noise limits the sensitivity of the detector to $\delta \nphoton \gtrsim 0.3$.
These two orders of magnitude constitute a suitable dynamic range for the optimizations that follow.

Photon depletion by feedback applies one of two depletion pulses, $D_j$, conditioned on the declared measurement result, $j\in \{0,1\}$ [Fig.~\ref{fig:fig2}(a)].
The pulse $D_j$, a square pulse of duration $\taup=30~\ns$, is applied at $f_{\mathrm{r},\ket{j}}$ by sideband modulating $\frf$.
The combined delays from round-trip signal propagation ($80~\ns$), the augmented integration window ($100~\ns$), and controller latency ($150~\ns$) make $D_j$ arrive $330~\ns$ after the measurement pulse ends.
The amplitude and phase of each pulse is separately optimized using a two-step procedure.
Using the modified AllXY sequence with the qubit initialized in $\ket{i}$, we first minimize $\nphoton$ at $\taud=1000~\ns$.
This $\taud$ is sufficiently long to avoid saturating the detector and to reach the sensitivity limit after a few optimization rounds.
Next, we minimize $\nphoton$ at $\taud=500~\ns$.
This second optimization converges to $\nphoton\sim 2.1~(0.7)$ for $\ket{0}$ ($\ket{1}$), reducing $\taud$ by at least $5/\kappa$ compared to passive depletion [Fig.~\ref{fig:fig2}(b)].
An incorrect assignment by the feedback controller leads to less effective depletion but still outperforms passive depletion.
We have also explored conditional depletion for various pulse lengths while fixing $\taud=500~\ns$.
We observe a systematic evolution of the optimal depletion pulse parameters but no further reduction of $\nphoton$~\cite{SOMdepletion}.

\begin{figure}[!htb]
  \centering
  \includegraphics[width=\columnwidth]{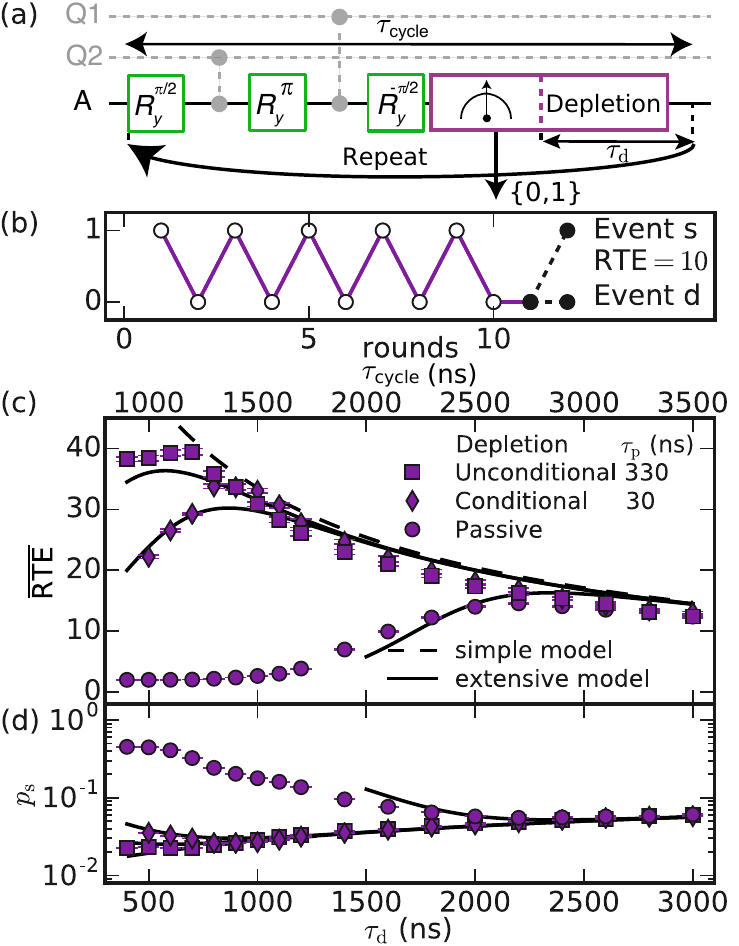}
  \caption{(Color online) {\bf Emulated multi-round QEC: flipping ancilla qubit.}
  (a) Block diagram for repeating parity measurements in a repetition code. The ancilla $A$ performs an indirect measurement of the parity of data qubits $\Qone$ and $\Qtwo$ by a coherent $200~\ns$ interaction step followed by measurement.
  This emulation replaces the c-phase gates by idling, reducing the coherent step to a simple echo sequence that ideally flips the ancilla each round.
  The measurement step is followed by a depletion step of duration $\taud$, after which a new cycle begins.
  (b) Single trace of digitized measurement outcomes. An event is detected whenever the measurement outcome first deviates from the ideal alternating sequence.
  Two types of event, $\types$ and $\typed$, are distinguished by the measurement outcome on the next round.
  (c) Average rounds to event as a function of $\taud$.
  The unconditional method improves $\RTEmean$ by a factor $2.7$ and reduces the optimum $\taud$ by a factor $>2.7$.
  (d) Per-round probability of type-$\types$ event versus $\taud$.
  Added curves are obtained from the two models described in the text.}
  \label{fig:fig3}
\end{figure}

Unconditional depletion uses a universal depletion pulse $\Dall$ starting immediately after the measurement pulse (there is no latency cost) [Fig.~\ref{fig:fig2}(c)].
This pulse is composed by summing two square pulses of duration $\taup=330~\ns$ with independent amplitude and phase, generated by sideband modulating $\frf$ at $\frzero$ and $\frone$.
These four parameters are numerically optimized using the sum of $\nphoton$ for $\ket{0}$ and $\ket{1}$ as cost function and a similar two-step procedure (with $\taud=400~\ns$ in the second step) as for the conditional pulses.
The minimization achieves $\nphoton\sim0.8~(0.4)$ for $\ket{0}$ ($\ket{1}$) and reduces $\taud$ by more than $6/\kappa$ compared to passive depletion [Fig.~\ref{fig:fig2}(d)].
We have also explored unconditional depletion for various pulse lengths while fixing $\taud=400~\ns$~\cite{SOMdepletion}.
We find a smooth variation of optimal pulse parameters, and a small improvement in residual $\nphoton$ with $\taup=270~\ns$. However, the overlap of the depletion pulse with the measurement integration window reduces the readout fidelity at this setting. 

We quantify the merits of these active photon depletion schemes with an experiment motivated by current efforts in multi-round quantum error correction~(QEC).
Specifically, we emulate an ancilla qubit undergoing the rapid succession of interleaved coherent interaction and measurement steps when performing repetitive parity checks on data-carrying qubits in a repetition code.
We replace each conditional-phase (c-phase) gate in the interaction step with idling for an equivalent time $(40~\ns)$, reducing the coherent step to a $200~\ns$ echo sequence that ideally flips the ancilla each round [Fig.~\ref{fig:fig3}(a)].
As performance metric, we measure the average number of rounds to an error detection event, $\RTEmean$.
An error event is marked by the first deviation of qubit measurement results from the ideal alternating sequence.
Imperfections reducing $\RTEmean$ include qubit relaxation, dephasing and detuning during the interaction step, and measurement errors due to readout discrimination infidelity, $1-\Fd$ (defined as the overlap fraction of gaussian best fits to the single-shot readout histograms~\cite{Sank14}).
To differentiate these sources of ancilla hardware errors, we keep track of two types of detection events, determined by the measurement outcome in the round following the first deviation (Fig.~\ref{fig:fig3}(b), similar to Ref.~\cite{Fowler14}).
Events of type $\types$ can result, for example, from a single ancilla bit flip or from measurement errors in two consecutive rounds.
In turn, events of type $\typed$ can result from one measurement error or from ancilla bit flips in two consecutive rounds.
Because photon-induced errors primarily lead to single ancilla bit flips, we also extract the probability of encountering an event of type $\types$ per cycle, $\psingle$, and investigate its $\taud$ dependence.

Decreasing $\taud$ trades off $\Tone$-induced errors for photon-induced errors.
For passive depletion, $\RTEmean$ is maximized to $14.6$ at $\taud=2200~\ns$ [Fig.~\ref{fig:fig3}(c)].
At this optimal point, depletion occupies most of the total QEC cycle time $\taucycle=2700~\ns$.
The active depletion methods reach a higher $\RTEmean$ by balancing the trade-off at lower $\taud$.
As in the optimization, we find that unconditional depletion performs best, improving the maximal $\RTEmean$ to $39.5$ and reducing the optimum $\taucycle$ to $1200~\ns$.

The essential features of $\RTEmean$ for the three depletion schemes are well captured by two theory models (see~\cite{SOMdepletion}).
The simple model includes only qubit relaxation and non-photon-induced dephasing (calibrated using standard $\Tone$ and $\Techo$ measurements).
The extensive model also includes photon-induced qubit dephasing and detuning during the idling steps (modeled following Ref.~\cite{FriskKockum12} with photon dynamics of Fig.~\ref{fig:fig2}), and a measured $1-\Fd=0.1\%$ for readout.
As we do not model photon-induced pulse errors, we restrict the extensive model to $\nphoton<8$.
The good agreement observed between the extensive model and experiment demonstrates the non-demolition character of the measurement and confirms the $\nphoton$ calibration.

\begin{figure}
    \centering
     \includegraphics[width=1\linewidth]{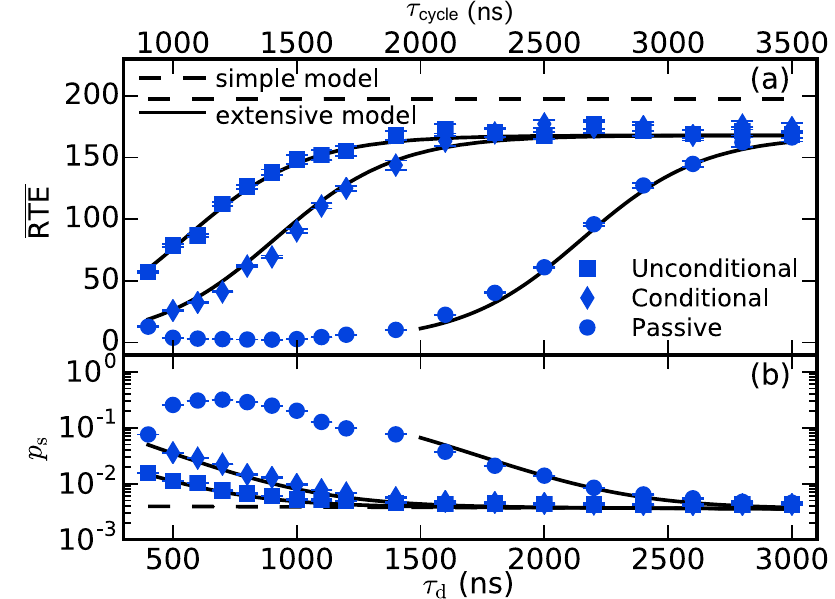}
     \caption{(Color online) {\bf Emulated multi-round QEC: non-flipping ancilla qubit in $\ket{0}$.}
     This variant uses the sequence of Fig.~\ref{fig:fig3}(a) but with opposite polarity on the final $\pi/2$ pulse in order not to flip the ancilla in every round.
     (a) Average rounds to event as a function of $\taud$, for ancilla starting in $\ket{0}$. $\RTEmean$ is no longer sensitive to qubit relaxation during $\taud$, increasing the sensitivity to $\nphoton$. The ceiling of $\sim 168$ reached at long $\taud$ is set by intrinsic decoherence in the coherent step and readout discrimination infidelity.
     (b) Per-round probability of encountering event of type $\types$ as a function of $\taud$. The simple and extensive models include the same calibrated errors as in Fig.~\ref{fig:fig3}.}
     \label{fig:fig4}
\end{figure}

The multi-round QEC emulation can be made more sensitive to leftover photons by harnessing the asymmetry of the qubit decay channel.
Specifically, changing the polarity of the final $\pi/2$ pulse so that the coherent step does not flip the ancilla can keep the ancilla ideally in $\ket{0}$ during measurement and depletion.
This change extends the sensitivity of $\RTEmean$ to $\nphoton$ by extending its ceiling to 168 [Fig.~\ref{fig:fig4}], which is $\taud$ independent and set by intrinsic decoherence in the coherent step and readout discrimination infidelity.
Clearly, unconditional depletion outperforms conditional and passive depletion, but the reduction of $\RTEmean$ to 50 at short $\taud$ evidences the performance limit reached by our choice of pulses.
In a QEC context, the key benefit of active depletion in this non-flipping variant will be an increase in $\RTEmean$ due to lower per-cycle probability of data qubit errors, afforded by reducing $\taucycle$ by $6/\kappa$. Evidently, this effect is not captured by our emulation, which is only sensitive to ancilla hardware errors.

These $\RTE$ experiments motivate two points for discussion and outlook.
First, they highlight the importance of using digital feedback~\cite{Riste12b} in QEC to keep ancillas in the ground state as much as possible (as used in a cat code~\cite{Ofek16}). Conveniently, this feedback has relaxed latency requirements ($\taud + 160~\ns$ in our example), because the conditional action can be chosen to be the polarity of the final $\pi/2$ pulse. Second, $\RTEmean$ emerges as an attractive method for quantifying the performance of every element in the QEC cycle, not just the depletion step. The advantage over traditional tune-up methods is the speed gained by not reinitializing in $\ket{0}$ following every measurement~\cite{Rol16} and the ability to tune without interrupting ongoing error correction~\cite{Kelly16}.

In summary, we have investigated two active methods for fast photon depletion in the nonlinear regime of cQED, relying on numerical optimization to successfully outperform passive depletion by more than $6/\kappa$.
Active photon depletion will find application in quantum computing scenarios interleaving qubit measurements with coherent qubit operations. Here, we have focused on the example of quantum error correction, emulating an ancilla qubit performing repetitive parity checks in a repetition code.
Future experiments will focus on combining active depletion with Purcell filtering to further reduce QEC cycle time from the achieved $1~\us$ to $\sim500~\ns$, sufficient to cross the error pseudo-threshold in small surface codes at state-of-the-art transmon relaxation times~\cite{Tomita14}.

\begin{acknowledgments}
We thank S.~Visser, J.~Somers, L.~Riesebos, and E.~Garrido~Barrab\'{e}s for contributions to FPGA programming, K.~W.~Lehnert for the parametric amplifier, A.~Wallraff for a precision coil, and K.~Bertels, E.~Charbon, D.~Sank, and M.~Khezri for discussions. We acknowledge funding from the EU FP7 project ScaleQIT, the Dutch Organization for Fundamental Research on Matter (FOM), an ERC Synergy Grant, a Marie Curie Career Integration Grant (L.D.C.), and the China Scholarship Council (X.F.).
\end{acknowledgments}


%

\clearpage

\renewcommand{\theequation}{S\arabic{equation}}
\renewcommand{\thefigure}{S\arabic{figure}}
\renewcommand{\thetable}{S\arabic{table}}
\setcounter{figure}{0}
\setcounter{equation}{0}
\setcounter{table}{0}{}

\section*{Supplementary material for  ``Active resonator reset in the nonlinear dispersive regime of circuit QED''}
\



This supplement provides additional figures and a description of the theoretical models used to model the QEC emulation experiments. 

\section{Additional Figures}
\label{sec:som_figs}
\subsection{Experimental setup}
\begin{figure*}[hbt!]
    \centering
     \includegraphics[width=1.4\columnwidth]{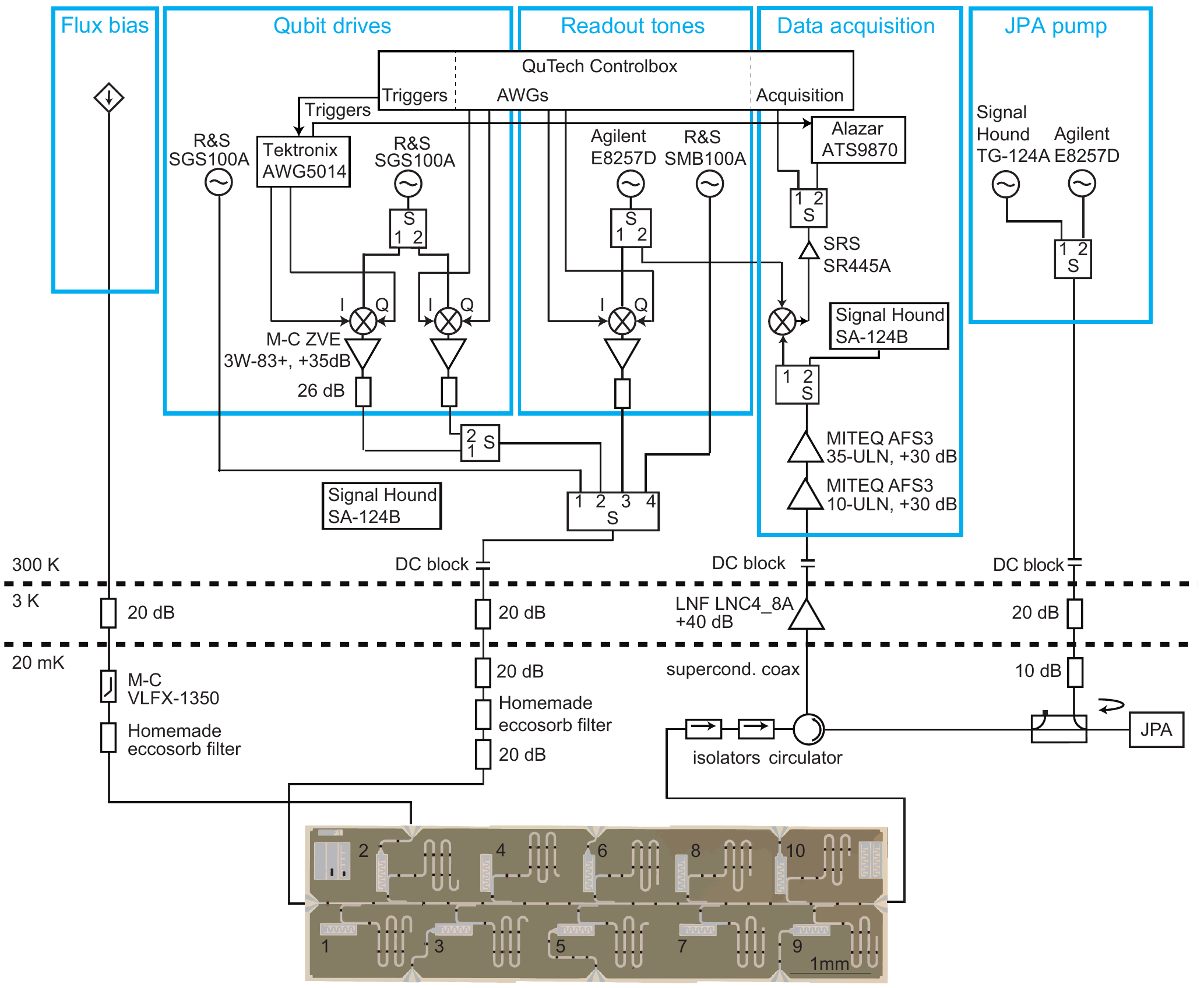}
     \caption{{\bf Experimental setup.} Photograph of the cQED chip and complete wiring diagram of electronic components inside and outside the $^3$He/$^4$He dilution refrigerator (Leiden Cryogenics CF-450). 
     The chip contains ten transmon qubits individually coupled to dedicated readout resonators.
     All resonators couple capacitively to the common feedline traversing the chip. 
     All data shown correspond to qubit-resonator pair 2. Dark features traversing the coplanar waveguide transmission lines are NbTiN bridges which interconnect ground planes and suppress slot-line mode propagation.
     }
     \label{fig:figS1}
\end{figure*}

Figure~\ref{fig:figS1} shows the device and experimental setup, including a full wiring diagram. The chip contains ten transmon qubit-resonator pairs. 
All experiments presented target pair 2.
The experimental setup is similar to that of previous experiments~\cite{Riste15_b}, but with an important addition labeled QuTech Control Box.
This homebuilt controller, comprised of 4 interconnected field-programmable gate arrays (Altera Cyclone IV), has digitizing and waveform generation capabilities.
The 2-channel digitizer samples with 8-bit resolution at $200~\MSps$. The 6-channel waveform generator produces qubit and resonator pulse envelopes with 14-bit resolution at $200~\MSps$.

\subsection{Photon number calibration}
\begin{figure}
      \centering
      \includegraphics[width=\linewidth]{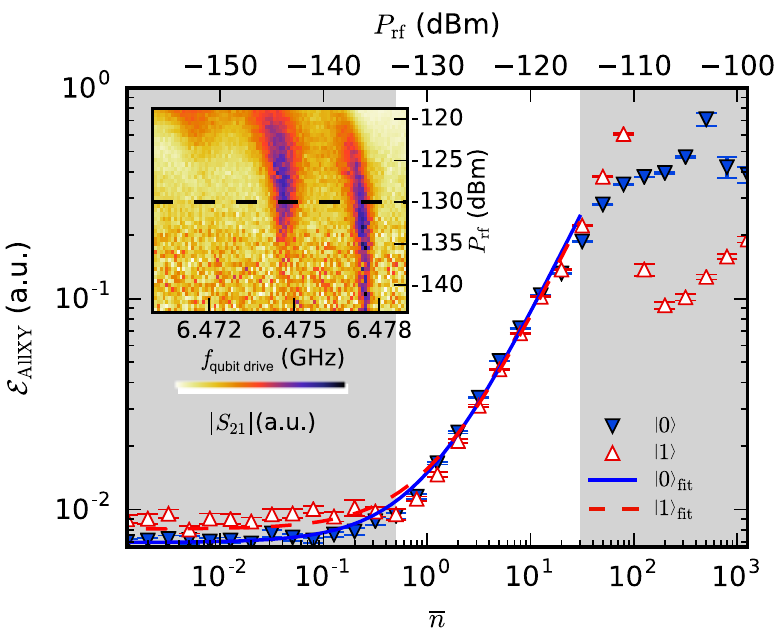}
      \caption{{\bf Calibration of photon number using AllXY error.}
      $\AllXYerror$ measured directly after a readout pulse of $1800~\ns$ duration drives the resonator into a steady-state photon population, $\nphoton$, for input states $\ket{0}$ and $\ket{1}$.
      The lines show a bilinear fit to the form $\AllXYerror=\alpha\nphoton+\beta$.
      Inset: photon-number splitting experiment~\cite{Schuster07_b} used to calibrate the single-photon power level, $\Prf\sim-130~\dBm$.}
      \label{fig:figS2}
\end{figure}

Figure~\ref{fig:figS2} contains the calibration of the photon number using AllXY error ($\AllXYerror$) as a detector.
$\AllXYerror$ is defined as the average absolute deviation from the ideal 2-step result in an AllXY experiment.
To calibrate the detector the resonator is populated using a long ($1800~\ns$) readout pulse with a varying pulse amplitude before measuring the AllXY.
This pulse amplitude is converted to an average photon number using the single-photon power that is extracted from a photon number splitting experiment.
We fit the form $\AllXYerror=\alpha\nphoton+\beta$ to the data for each input state separately, with $\alpha$ and $\beta$ as free parameters.
The best-fit functions are used throughout the experiment to convert $\AllXYerror$ to $\nphoton$.

\subsection{Constant excited state QEC emulation}
\begin{figure}
    \centering
     \includegraphics[width=1\linewidth]{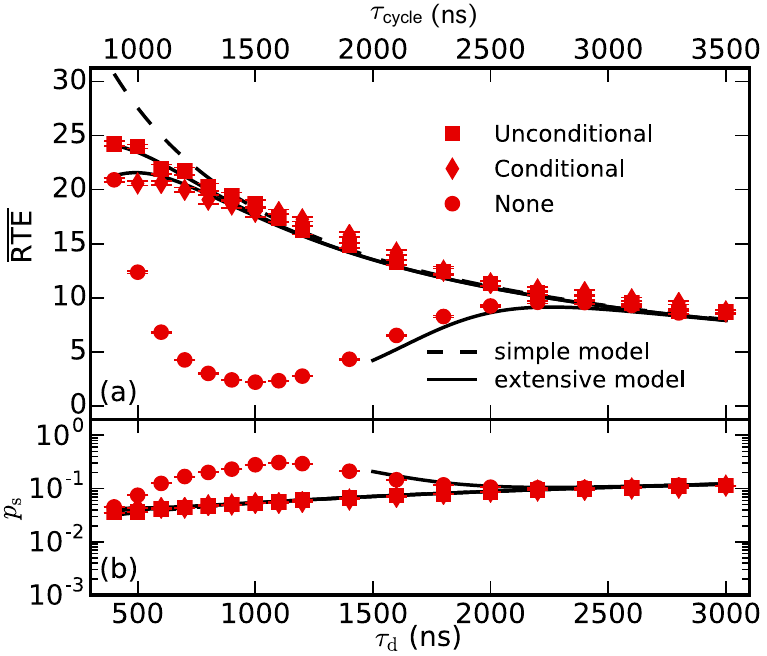}
     \caption{{\bf Emulated multi-round QEC: non-flipping ancilla in $\ket{1}$.}
     This variant of the emulation uses the same sequence as Fig.~4 but with the qubit initialized in $\ket{1}$.
     (a) Mean rounds to error detection event, $\RTEmean$, as a function of $\taud$.
     (b) Per-round probability of encountering event of type $\types$ as a function of $\taud$.
     Added curves correspond to the simple and extensive models described in Sec.~\ref{sec:model}.}
     \label{fig:figS3}
\end{figure}

Figure~\ref{fig:figS3} shows the emulated multi-round QEC for a non-flipping ancilla when the qubit is initialized in the excited state.
This variant of the emulation uses the same sequence as Fig.~4 but with the qubit initialized in $\ket{1}$. 
Varying $\taud$, we find the optimum tradeoff between errors induced by leftover photons and by relaxation for the three methods. Unconditional depletions performs best, increasing $\RTEmean$ by a factor $2.5$ with respect to passive depletion. Note that passive depletion produces a spurious increase in $\RTEmean$ for very short $\taud$. 
The high photon number detunes the qubit so much that qubit pulses are inoperative, causing the qubit to remain in the same state and yielding long strings of identical, expected measurement outcomes.

\subsection{Optimal depletion pulse characterization}
Figures~S4 and~S5 summarize our further investigation of depletion-pulse optimizations for conditional and unconditional depletion, respectively.
For a variety of pulse lengths $\taup$, the optimized pulse amplitudes and phase parameters are shown, along with residual photon levels and results for multi-round QEC emulation.

For conditional depletion, the optimal amplitude $A_0$ ($A_1$) of $\Dzero$ ($\Done$) decreases smoothly as $\taup$ increases, whereas the optimal phase $\phi_0$ ($\phi_1$) remains constant.
The discrimination infidelity $1-\Fd$ is inferred from single-shot readout histogram experiments and is defined as the fraction of overlap of the best-fit gaussians.
The residual photon number and readout discrimination infidelity do not show any dependence on $\taup$.
As expected, there is no dependence of the fidelity on $\taup$ as there is no overlap between the depletion pulse and integration window.
The average rounds to event and per-round probability of type-$\types$ event for emulated QEC in the flipping configuration do not show any dependence on $\taup$ either.

For unconditional depletion, the optimal values of the four parameters defining the universal depletion pulse $\Dall$ evolve smoothly as $\taup$ is varied.
The optimized $\nphoton$ first decreases weakly with decreasing $\taup$ but increases sharply for $\taup<250~\ns$.
A smooth decrease in $\Fd$ is observed for decreasing $\taup$. We attribute this effect to the overlap between $\Dall$ and the measurement integration window.
$\RTEmean$ is unchanged for $\taup>270~\ns$, suggesting a trade-off between errors due to $\nphoton$ and $\Fd$.
This trade-off is reflected in the corresponding increase of per-round probability of type-$\types$ event.

\begin{figure}
    \centering
    \includegraphics[width=1\linewidth]{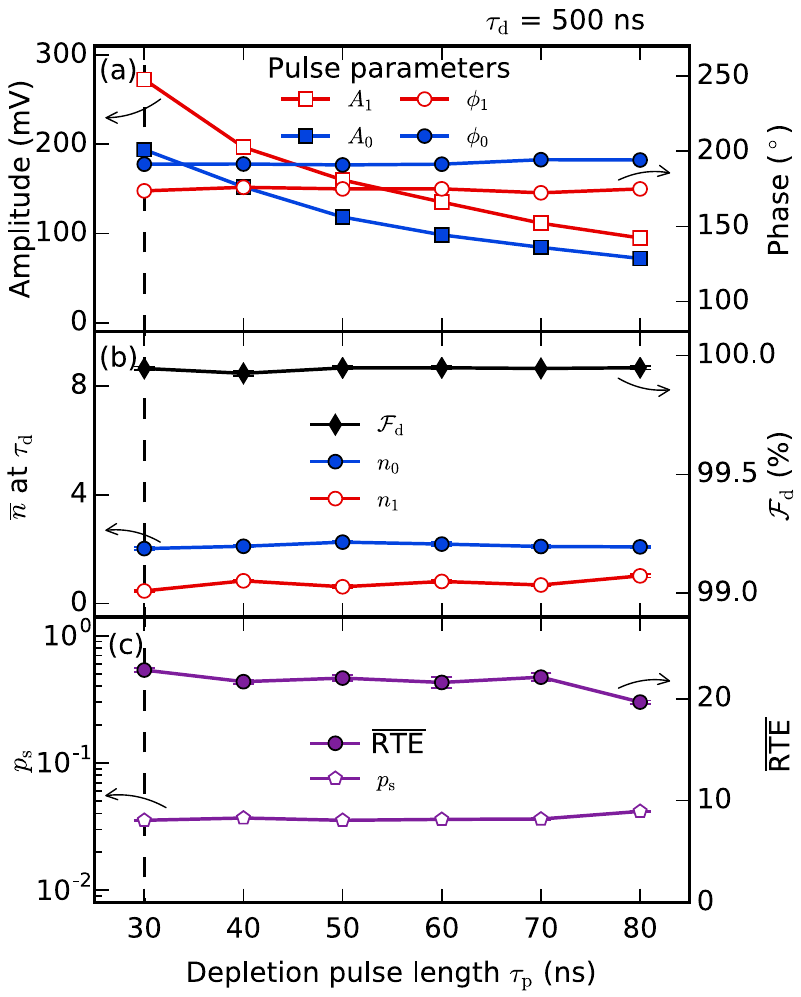}
    \caption{{\bf Characterization of conditional depletion as a function of depletion pulse length $\taup$.}
    The dashed line indicates $\taup=30~\ns$, used in Figs.~2 to 4 and Fig.~\ref{fig:figS3}.
    All data were taken at a fixed $\taud=500~\ns$.
    (a) Optimal pulse parameters.
    (b) Residual photon number for both qubit states and discrimination fidelity $\Fd$.
    (c) Average rounds to event and per-round probability of type-$\types$ event for emulated QEC in the flipping configuration.
}
     \label{fig:figS4}
\end{figure}

\begin{figure}
  \centering
  \includegraphics[width=1\linewidth]{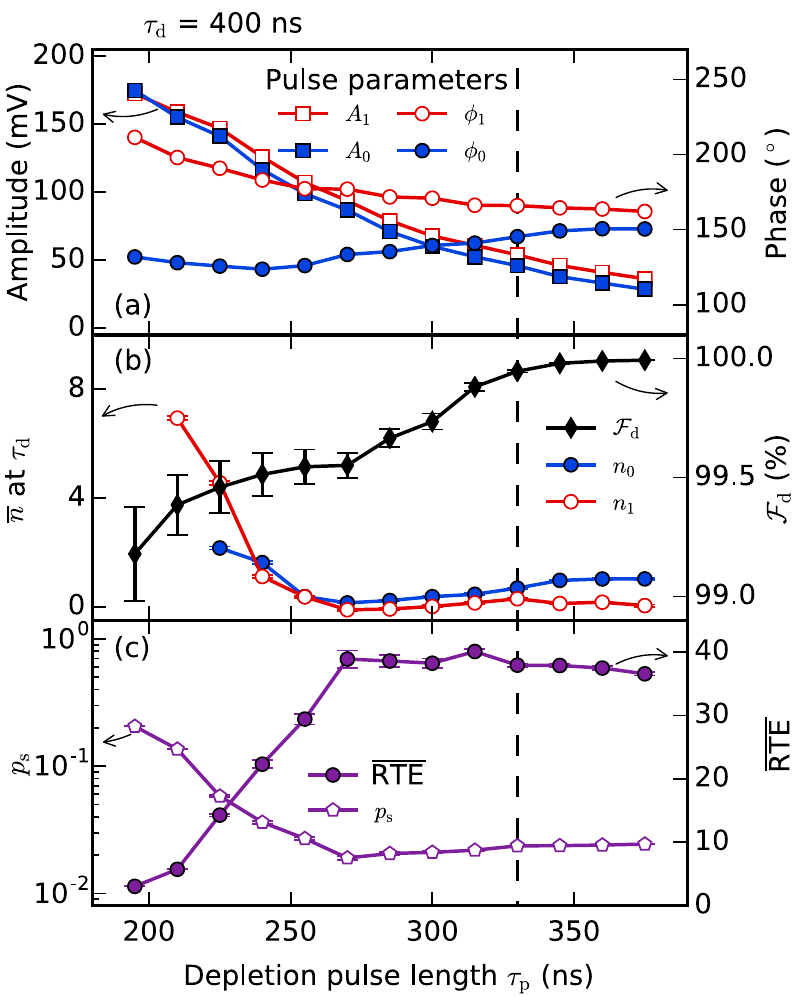}
  \caption{{\bf Characterization of unconditional depletion as a function of depletion pulse length $\taup$.}
    The dashed line indicates $\taup=330~\ns$, used in Figs.~2 to 4 and Fig.~\ref{fig:figS3}.
    All data were taken at a fixed depletion time of $\taud=400~\ns$
    (a) Optimal pulse parameters.
    (b) Residual photon number for both qubit states and discrimination fidelity $\Fd$.
    (c) Average rounds to event and per-round probability of $\types$-event for emulated QEC in the flipping configuration.
}
  \label{fig:figS5}
\end{figure}

\newpage
\section{Theoretical Models}
\label{sec:model}
We use two models to compare to data in Figs.~3,~4, and~S3 labelled simple and extensive.
The simple model includes ancilla relaxation and intrinsic dephasing, providing an upper bound for the performance of the emulated multi-round QEC circuit.
The extensive model further includes ancilla readout error and detuning and dephasing from the photon-induced AC Stark shift.
These models use separately calibrated parameters.

The ancilla sans photon field is modeled considering amplitude and phase damping as in~\cite{Nielsen00}. 
Single-qubit gates are approximated as $40~\ns$ decay windows with perfect instantaneous pulses in the middle.
This leads to the following scheme: $\taud+20~\ns$ of $\Tone$ decay, followed by a $\pi/2$ pulse, then $160~\ns$ of $\Techo$ decay (with a $\pi$ pulse in the middle), another $\pi/2$ pulse, and $20~\ns$ of $\Tone$ decay.

Measurement is modeled as a perfect state update $S_1$, followed by a $\taumeas=300~\ns$ decay window, and a second state update $S_2$. The measurement signal is conditioned both on the state post-$S_1$ ($|\psi_i\>$) and post-$S_2$ ($|\psi_o\>$). If $|\psi_i\>=|\psi_o\>$ no decay occurred, and the incorrect measurement is returned with probability $1-\Fd=0.1\%$ [Fig.~S4(b)].
The only other possibility is for a single decay event (as we do not allow excitations). To zeroth order in $\taumeas/\Tone\approx 1/800$, this situation has equal probability of returning either measurement signal.

During the coherent phase, the off-diagonal elements are affected by the photon population. We model this effect following Ref.~\onlinecite{FriskKockum12_b}:
\begin{equation}
\frac{d\rhoqb}{dt}=-i\frac{\bar{\omega}_\mathrm{a}+B}{2}[\sigma_\mathrm{z},\rhoqb]+\gamma_1\Dd[\sigma_-]\rhoqb+\frac{\gamma_\mathrm{\phi}+\Gamma_\mathrm{d}}{2}\Dd[\sigma_\mathrm{z}]\rhoqb.
\end{equation}
Here, $\Dd[X]$ is the Lindblad operator $\Dd[X]\rho=X\rho X^{\dag}-\frac{1}{2}X^\dag X\rho-\frac{1}{2}\rho X^\dag X$, $\gamma_1=1/\Tone$ and $\gamma_\mathrm{\phi}$ the pure dephasing rate [$\gamma_\mathrm{\phi}=(\Techo)^{-1}-\frac{1}{2}\Tone^{-1}=(177\us)^{-1}$].
$\bar{\omega}_\mathrm{a}$ is a constant rotation around the $z$ axis of the Bloch sphere, and so is canceled by the $\pi$ pulse in the coherent phase.
$\Gamma_\mathrm{d}=2\chi\text{Im}(\alpha_0\alpha_1^*)$ is the measurement-induced dephasing, with $\alpha_{0,1}$ the qubit-state-dependent photon field amplitude and $2\chi$ the dispersive shift per photon.
This contributes a decay to the off-diagonal element of the density matrix during the coherent phase, multiplying it by
\begin{equation}
\exp\left[-\int\Gamma_\mathrm{d}(t)\right],
\end{equation}
where the integral is taken over the coherent time window. $B=2\chi\text{Re}(\alpha_0\alpha_1^*)$ is the AC Stark shift, which detunes the ancilla by an amount equal to the difference in the average photon number over the two parts of the coherent phase. This multiplies the off-diagonal terms by a complex phase
\begin{equation}
\phi_\mathrm{Stark}=\int_{t_\mathrm{A}}B(t)-\int_{t_\mathrm{B}}B(t).
\end{equation}
Here, $t_A$ and $t_B$ are the time windows in the coherent phase on either side of the $\pi$ pulse.
The magnitude of the photon fields post-depletion is taken from Fig.~2, and experiences an exponential decay at a rate that is obtained by fitting curves to the same figure.
The phase difference between the fields associated with the ground and excited state grows at a rate $2\chi$, as extracted from Fig.~1.
As we do not model photon-induced pulse errors, we restrict our modeling to $\nphoton<8$, where these effects are negligible.

The experiment is simulated by storing the error-free ancilla population as a unnormalized density matrix and applying repeated cycles of the circuit.
At each measurement step, the fraction of the density matrix that corresponded to an event is removed and the corresponding probability stored.
The removed fraction of the density matrix in evolved for one more cycle in order to extract the event type probabilities.
This is repeated until the remaining population is less than $10^{-6}$.


\begin{thebibliography}{37}%
\makeatletter
\providecommand \@ifxundefined [1]{%
 \@ifx{#1\undefined}
}%
\providecommand \@ifnum [1]{%
 \ifnum #1\expandafter \@firstoftwo
 \else \expandafter \@secondoftwo
 \fi
}%
\providecommand \@ifx [1]{%
 \ifx #1\expandafter \@firstoftwo
 \else \expandafter \@secondoftwo
 \fi
}%
\providecommand \natexlab [1]{#1}%
\providecommand \enquote  [1]{``#1''}%
\providecommand \bibnamefont  [1]{#1}%
\providecommand \bibfnamefont [1]{#1}%
\providecommand \citenamefont [1]{#1}%
\providecommand \href@noop [0]{\@secondoftwo}%
\providecommand \href [0]{\begingroup \@sanitize@url \@href}%
\providecommand \@href[1]{\@@startlink{#1}\@@href}%
\providecommand \@@href[1]{\endgroup#1\@@endlink}%
\providecommand \@sanitize@url [0]{\catcode `\\12\catcode `\$12\catcode
  `\&12\catcode `\#12\catcode `\^12\catcode `\_12\catcode `\%12\relax}%
\providecommand \@@startlink[1]{}%
\providecommand \@@endlink[0]{}%
\providecommand \url  [0]{\begingroup\@sanitize@url \@url }%
\providecommand \@url [1]{\endgroup\@href {#1}{\urlprefix }}%
\providecommand \urlprefix  [0]{URL }%
\providecommand \Eprint [0]{\href }%
\providecommand \doibase [0]{http://dx.doi.org/}%
\providecommand \selectlanguage [0]{\@gobble}%
\providecommand \bibinfo  [0]{\@secondoftwo}%
\providecommand \bibfield  [0]{\@secondoftwo}%
\providecommand \translation [1]{[#1]}%
\providecommand \BibitemOpen [0]{}%
\providecommand \bibitemStop [0]{}%
\providecommand \bibitemNoStop [0]{.\EOS\space}%
\providecommand \EOS [0]{\spacefactor3000\relax}%
\providecommand \BibitemShut  [1]{\csname bibitem#1\endcsname}%
\let\auto@bib@innerbib\@empty
\bibitem [{\citenamefont {Reed}\ \emph {et~al.}(2012)\citenamefont {Reed},
  \citenamefont {DiCarlo}, \citenamefont {Nigg}, \citenamefont {Sun},
  \citenamefont {Frunzio}, \citenamefont {Girvin},\ and\ \citenamefont
  {Schoelkopf}}]{Reed12}%
  \BibitemOpen
  \bibfield  {author} {\bibinfo {author} {\bibfnamefont {M.~D.}\ \bibnamefont
  {Reed}}, \bibinfo {author} {\bibfnamefont {L.}~\bibnamefont {DiCarlo}},
  \bibinfo {author} {\bibfnamefont {S.~E.}\ \bibnamefont {Nigg}}, \bibinfo
  {author} {\bibfnamefont {L.}~\bibnamefont {Sun}}, \bibinfo {author}
  {\bibfnamefont {L.}~\bibnamefont {Frunzio}}, \bibinfo {author} {\bibfnamefont
  {S.~M.}\ \bibnamefont {Girvin}}, \ and\ \bibinfo {author} {\bibfnamefont
  {R.~J.}\ \bibnamefont {Schoelkopf}},\ }\href@noop {} {\bibfield  {journal}
  {\bibinfo  {journal} {Nature}\ }\textbf {\bibinfo {volume} {482}},\ \bibinfo
  {pages} {382} (\bibinfo {year} {2012})}\BibitemShut {NoStop}%
\bibitem [{\citenamefont {Kelly}\ \emph {et~al.}(2015)\citenamefont {Kelly},
  \citenamefont {Barends}, \citenamefont {Fowler}, \citenamefont {Megrant},
  \citenamefont {Jeffrey}, \citenamefont {White}, \citenamefont {Sank},
  \citenamefont {Mutus}, \citenamefont {Campbell}, \citenamefont {Chen} \emph
  {et~al.}}]{Kelly15}%
  \BibitemOpen
  \bibfield  {author} {\bibinfo {author} {\bibfnamefont {J.}~\bibnamefont
  {Kelly}}, \bibinfo {author} {\bibfnamefont {R.}~\bibnamefont {Barends}},
  \bibinfo {author} {\bibfnamefont {A.}~\bibnamefont {Fowler}}, \bibinfo
  {author} {\bibfnamefont {A.}~\bibnamefont {Megrant}}, \bibinfo {author}
  {\bibfnamefont {E.}~\bibnamefont {Jeffrey}}, \bibinfo {author} {\bibfnamefont
  {T.}~\bibnamefont {White}}, \bibinfo {author} {\bibfnamefont
  {D.}~\bibnamefont {Sank}}, \bibinfo {author} {\bibfnamefont {J.}~\bibnamefont
  {Mutus}}, \bibinfo {author} {\bibfnamefont {B.}~\bibnamefont {Campbell}},
  \bibinfo {author} {\bibfnamefont {Y.}~\bibnamefont {Chen}},  \emph {et~al.},\
  }\href@noop {} {\bibfield  {journal} {\bibinfo  {journal} {Nature}\ }\textbf
  {\bibinfo {volume} {519}},\ \bibinfo {pages} {66} (\bibinfo {year}
  {2015})}\BibitemShut {NoStop}%
\bibitem [{\citenamefont {Rist\`{e}}\ \emph {et~al.}(2015)\citenamefont
  {Rist\`{e}}, \citenamefont {Poletto}, \citenamefont {Huang}, \citenamefont
  {Bruno}, \citenamefont {Vesterinen}, \citenamefont {Saira},\ and\
  \citenamefont {DiCarlo}}]{Riste15}%
  \BibitemOpen
  \bibfield  {author} {\bibinfo {author} {\bibfnamefont {D.}~\bibnamefont
  {Rist\`{e}}}, \bibinfo {author} {\bibfnamefont {S.}~\bibnamefont {Poletto}},
  \bibinfo {author} {\bibfnamefont {M.~Z.}\ \bibnamefont {Huang}}, \bibinfo
  {author} {\bibfnamefont {A.}~\bibnamefont {Bruno}}, \bibinfo {author}
  {\bibfnamefont {V.}~\bibnamefont {Vesterinen}}, \bibinfo {author}
  {\bibfnamefont {O.~P.}\ \bibnamefont {Saira}}, \ and\ \bibinfo {author}
  {\bibfnamefont {L.}~\bibnamefont {DiCarlo}},\ }\href@noop {} {\bibfield
  {journal} {\bibinfo  {journal} {Nat.\ Commun.}\ }\textbf {\bibinfo {volume}
  {{6}}} (\bibinfo {year} {{2015}})}\BibitemShut {NoStop}%
\bibitem [{\citenamefont {Cramer}\ \emph {et~al.}(2015)\citenamefont {Cramer},
  \citenamefont {Kalb}, \citenamefont {Rol}, \citenamefont {Hensen},
  \citenamefont {Markham}, \citenamefont {Twitchen}, \citenamefont {Hanson},\
  and\ \citenamefont {Taminiau}}]{Cramer15}%
  \BibitemOpen
  \bibfield  {author} {\bibinfo {author} {\bibfnamefont {J.}~\bibnamefont
  {Cramer}}, \bibinfo {author} {\bibfnamefont {N.}~\bibnamefont {Kalb}},
  \bibinfo {author} {\bibfnamefont {M.~A.}\ \bibnamefont {Rol}}, \bibinfo
  {author} {\bibfnamefont {B.}~\bibnamefont {Hensen}}, \bibinfo {author}
  {\bibfnamefont {M.}~\bibnamefont {Markham}}, \bibinfo {author} {\bibfnamefont
  {D.~J.}\ \bibnamefont {Twitchen}}, \bibinfo {author} {\bibfnamefont
  {R.}~\bibnamefont {Hanson}}, \ and\ \bibinfo {author} {\bibfnamefont {T.~H.}\
  \bibnamefont {Taminiau}},\ }\href@noop {} {\bibfield  {journal} {\bibinfo
  {journal} {arXiv:1508.01388v1}\ } (\bibinfo {year} {2015})}\BibitemShut
  {NoStop}%
\bibitem [{\citenamefont {Nigg}\ \emph {et~al.}(2014)\citenamefont {Nigg},
  \citenamefont {M\"{u}ller}, \citenamefont {Martinez}, \citenamefont
  {Schindler}, \citenamefont {Hennrich}, \citenamefont {Monz}, \citenamefont
  {Martin-Delgado},\ and\ \citenamefont {Blatt}}]{Nigg14}%
  \BibitemOpen
  \bibfield  {author} {\bibinfo {author} {\bibfnamefont {D.}~\bibnamefont
  {Nigg}}, \bibinfo {author} {\bibfnamefont {M.}~\bibnamefont {M\"{u}ller}},
  \bibinfo {author} {\bibfnamefont {E.~A.}\ \bibnamefont {Martinez}}, \bibinfo
  {author} {\bibfnamefont {P.}~\bibnamefont {Schindler}}, \bibinfo {author}
  {\bibfnamefont {M.}~\bibnamefont {Hennrich}}, \bibinfo {author}
  {\bibfnamefont {T.}~\bibnamefont {Monz}}, \bibinfo {author} {\bibfnamefont
  {M.~A.}\ \bibnamefont {Martin-Delgado}}, \ and\ \bibinfo {author}
  {\bibfnamefont {R.}~\bibnamefont {Blatt}},\ }\href@noop {} {\bibfield
  {journal} {\bibinfo  {journal} {Science}\ }\textbf {\bibinfo {volume}
  {345}},\ \bibinfo {pages} {302} (\bibinfo {year} {2014})}\BibitemShut
  {NoStop}%
\bibitem [{\citenamefont {Corcoles}\ \emph {et~al.}(2015)\citenamefont
  {Corcoles}, \citenamefont {Magesan}, \citenamefont {Srinivasan},
  \citenamefont {Cross}, \citenamefont {Steffen}, \citenamefont {Gambetta},\
  and\ \citenamefont {Chow}}]{Corcoles15}%
  \BibitemOpen
  \bibfield  {author} {\bibinfo {author} {\bibfnamefont {A.~D.}\ \bibnamefont
  {Corcoles}}, \bibinfo {author} {\bibfnamefont {E.}~\bibnamefont {Magesan}},
  \bibinfo {author} {\bibfnamefont {S.~J.}\ \bibnamefont {Srinivasan}},
  \bibinfo {author} {\bibfnamefont {A.~W.}\ \bibnamefont {Cross}}, \bibinfo
  {author} {\bibfnamefont {M.}~\bibnamefont {Steffen}}, \bibinfo {author}
  {\bibfnamefont {J.~M.}\ \bibnamefont {Gambetta}}, \ and\ \bibinfo {author}
  {\bibfnamefont {J.~M.}\ \bibnamefont {Chow}},\ }\href@noop {} {\bibfield
  {journal} {\bibinfo  {journal} {Nat.\ Commun.}\ }\textbf {\bibinfo {volume}
  {{6}}} (\bibinfo {year} {{2015}})}\BibitemShut {NoStop}%
\bibitem [{\citenamefont {Ofek}\ \emph {et~al.}(2016)\citenamefont {Ofek},
  \citenamefont {Petrenko}, \citenamefont {Heeres}, \citenamefont {Reinhold},
  \citenamefont {Leghtas}, \citenamefont {Vlastakis}, \citenamefont {Liu},
  \citenamefont {Frunzio}, \citenamefont {Girvin}, \citenamefont {Jiang},
  \citenamefont {Mirrahimi}, \citenamefont {Devoret},\ and\ \citenamefont
  {Schoelkopf}}]{Ofek16}%
  \BibitemOpen
  \bibfield  {author} {\bibinfo {author} {\bibfnamefont {N.}~\bibnamefont
  {Ofek}}, \bibinfo {author} {\bibfnamefont {A.}~\bibnamefont {Petrenko}},
  \bibinfo {author} {\bibfnamefont {R.}~\bibnamefont {Heeres}}, \bibinfo
  {author} {\bibfnamefont {P.}~\bibnamefont {Reinhold}}, \bibinfo {author}
  {\bibfnamefont {Z.}~\bibnamefont {Leghtas}}, \bibinfo {author} {\bibfnamefont
  {B.}~\bibnamefont {Vlastakis}}, \bibinfo {author} {\bibfnamefont
  {Y.}~\bibnamefont {Liu}}, \bibinfo {author} {\bibfnamefont {L.}~\bibnamefont
  {Frunzio}}, \bibinfo {author} {\bibfnamefont {S.~M.}\ \bibnamefont {Girvin}},
  \bibinfo {author} {\bibfnamefont {L.}~\bibnamefont {Jiang}}, \bibinfo
  {author} {\bibfnamefont {M.}~\bibnamefont {Mirrahimi}}, \bibinfo {author}
  {\bibfnamefont {M.~H.}\ \bibnamefont {Devoret}}, \ and\ \bibinfo {author}
  {\bibfnamefont {R.~J.}\ \bibnamefont {Schoelkopf}},\ }\href@noop {}
  {\bibfield  {journal} {\bibinfo  {journal} {arXiv:1602.04768}\ } (\bibinfo
  {year} {2016})}\BibitemShut {NoStop}%
\bibitem [{\citenamefont {Blais}\ \emph {et~al.}(2004)\citenamefont {Blais},
  \citenamefont {Huang}, \citenamefont {Wallraff}, \citenamefont {Girvin},\
  and\ \citenamefont {Schoelkopf}}]{Blais04}%
  \BibitemOpen
  \bibfield  {author} {\bibinfo {author} {\bibfnamefont {A.}~\bibnamefont
  {Blais}}, \bibinfo {author} {\bibfnamefont {R.-S.}\ \bibnamefont {Huang}},
  \bibinfo {author} {\bibfnamefont {A.}~\bibnamefont {Wallraff}}, \bibinfo
  {author} {\bibfnamefont {S.~M.}\ \bibnamefont {Girvin}}, \ and\ \bibinfo
  {author} {\bibfnamefont {R.~J.}\ \bibnamefont {Schoelkopf}},\ }\href@noop {}
  {\bibfield  {journal} {\bibinfo  {journal} {Phys. Rev. A}\ }\textbf {\bibinfo
  {volume} {69}},\ \bibinfo {pages} {062320} (\bibinfo {year}
  {2004})}\BibitemShut {NoStop}%
\bibitem [{\citenamefont {Petersson}\ \emph {et~al.}(2012)\citenamefont
  {Petersson}, \citenamefont {McFaul}, \citenamefont {Schroer}, \citenamefont
  {Jung}, \citenamefont {Taylor}, \citenamefont {Houck},\ and\ \citenamefont
  {Petta}}]{Petersson12}%
  \BibitemOpen
  \bibfield  {author} {\bibinfo {author} {\bibfnamefont {K.~D.}\ \bibnamefont
  {Petersson}}, \bibinfo {author} {\bibfnamefont {L.~W.}\ \bibnamefont
  {McFaul}}, \bibinfo {author} {\bibfnamefont {M.~D.}\ \bibnamefont {Schroer}},
  \bibinfo {author} {\bibfnamefont {M.}~\bibnamefont {Jung}}, \bibinfo {author}
  {\bibfnamefont {J.~M.}\ \bibnamefont {Taylor}}, \bibinfo {author}
  {\bibfnamefont {A.~A.}\ \bibnamefont {Houck}}, \ and\ \bibinfo {author}
  {\bibfnamefont {J.~R.}\ \bibnamefont {Petta}},\ }\href
  {http://dx.doi.org/10.1038/nature11559} {\bibfield  {journal} {\bibinfo
  {journal} {Nature}\ }\textbf {\bibinfo {volume} {490}},\ \bibinfo {pages}
  {380} (\bibinfo {year} {2012})}\BibitemShut {NoStop}%
\bibitem [{\citenamefont {Larsen}\ \emph {et~al.}(2015)\citenamefont {Larsen},
  \citenamefont {Petersson}, \citenamefont {Kuemmeth}, \citenamefont
  {Jespersen}, \citenamefont {Krogstrup}, \citenamefont {Nyg\aa{}rd},\ and\
  \citenamefont {Marcus}}]{Larsen15}%
  \BibitemOpen
  \bibfield  {author} {\bibinfo {author} {\bibfnamefont {T.~W.}\ \bibnamefont
  {Larsen}}, \bibinfo {author} {\bibfnamefont {K.~D.}\ \bibnamefont
  {Petersson}}, \bibinfo {author} {\bibfnamefont {F.}~\bibnamefont {Kuemmeth}},
  \bibinfo {author} {\bibfnamefont {T.~S.}\ \bibnamefont {Jespersen}}, \bibinfo
  {author} {\bibfnamefont {P.}~\bibnamefont {Krogstrup}}, \bibinfo {author}
  {\bibfnamefont {J.}~\bibnamefont {Nyg\aa{}rd}}, \ and\ \bibinfo {author}
  {\bibfnamefont {C.~M.}\ \bibnamefont {Marcus}},\ }\href@noop {} {\bibfield
  {journal} {\bibinfo  {journal} {Phys. Rev. Lett.}\ }\textbf {\bibinfo
  {volume} {115}},\ \bibinfo {pages} {127001} (\bibinfo {year}
  {2015})}\BibitemShut {NoStop}%
\bibitem [{\citenamefont {de~Lange}\ \emph {et~al.}(2015)\citenamefont
  {de~Lange}, \citenamefont {van Heck}, \citenamefont {Bruno}, \citenamefont
  {van Woerkom}, \citenamefont {Geresdi}, \citenamefont {Plissard},
  \citenamefont {Bakkers}, \citenamefont {Akhmerov},\ and\ \citenamefont
  {DiCarlo}}]{deLange15}%
  \BibitemOpen
  \bibfield  {author} {\bibinfo {author} {\bibfnamefont {G.}~\bibnamefont
  {de~Lange}}, \bibinfo {author} {\bibfnamefont {B.}~\bibnamefont {van Heck}},
  \bibinfo {author} {\bibfnamefont {A.}~\bibnamefont {Bruno}}, \bibinfo
  {author} {\bibfnamefont {D.~J.}\ \bibnamefont {van Woerkom}}, \bibinfo
  {author} {\bibfnamefont {A.}~\bibnamefont {Geresdi}}, \bibinfo {author}
  {\bibfnamefont {S.~R.}\ \bibnamefont {Plissard}}, \bibinfo {author}
  {\bibfnamefont {E.~P. A.~M.}\ \bibnamefont {Bakkers}}, \bibinfo {author}
  {\bibfnamefont {A.~R.}\ \bibnamefont {Akhmerov}}, \ and\ \bibinfo {author}
  {\bibfnamefont {L.}~\bibnamefont {DiCarlo}},\ }\href@noop {} {\bibfield
  {journal} {\bibinfo  {journal} {Phys. Rev. Lett.}\ }\textbf {\bibinfo
  {volume} {115}},\ \bibinfo {pages} {127002} (\bibinfo {year}
  {2015})}\BibitemShut {NoStop}%
\bibitem [{\citenamefont {Groen}\ \emph {et~al.}(2013)\citenamefont {Groen},
  \citenamefont {Rist\`e}, \citenamefont {Tornberg}, \citenamefont {Cramer},
  \citenamefont {de~Groot}, \citenamefont {Picot}, \citenamefont {Johansson},\
  and\ \citenamefont {DiCarlo}}]{Groen13}%
  \BibitemOpen
  \bibfield  {author} {\bibinfo {author} {\bibfnamefont {J.~P.}\ \bibnamefont
  {Groen}}, \bibinfo {author} {\bibfnamefont {D.}~\bibnamefont {Rist\`e}},
  \bibinfo {author} {\bibfnamefont {L.}~\bibnamefont {Tornberg}}, \bibinfo
  {author} {\bibfnamefont {J.}~\bibnamefont {Cramer}}, \bibinfo {author}
  {\bibfnamefont {P.~C.}\ \bibnamefont {de~Groot}}, \bibinfo {author}
  {\bibfnamefont {T.}~\bibnamefont {Picot}}, \bibinfo {author} {\bibfnamefont
  {G.}~\bibnamefont {Johansson}}, \ and\ \bibinfo {author} {\bibfnamefont
  {L.}~\bibnamefont {DiCarlo}},\ }\href@noop {} {\bibfield  {journal} {\bibinfo
   {journal} {Phys. Rev. Lett.}\ }\textbf {\bibinfo {volume} {111}},\ \bibinfo
  {pages} {090506} (\bibinfo {year} {2013})}\BibitemShut {NoStop}%
\bibitem [{\citenamefont {Johnson}\ \emph {et~al.}(2012)\citenamefont
  {Johnson}, \citenamefont {Macklin}, \citenamefont {Slichter}, \citenamefont
  {Vijay}, \citenamefont {Weingarten}, \citenamefont {Clarke},\ and\
  \citenamefont {Siddiqi}}]{Johnson12}%
  \BibitemOpen
  \bibfield  {author} {\bibinfo {author} {\bibfnamefont {J.~E.}\ \bibnamefont
  {Johnson}}, \bibinfo {author} {\bibfnamefont {C.}~\bibnamefont {Macklin}},
  \bibinfo {author} {\bibfnamefont {D.~H.}\ \bibnamefont {Slichter}}, \bibinfo
  {author} {\bibfnamefont {R.}~\bibnamefont {Vijay}}, \bibinfo {author}
  {\bibfnamefont {E.~B.}\ \bibnamefont {Weingarten}}, \bibinfo {author}
  {\bibfnamefont {J.}~\bibnamefont {Clarke}}, \ and\ \bibinfo {author}
  {\bibfnamefont {I.}~\bibnamefont {Siddiqi}},\ }\href {\doibase
  10.1103/PhysRevLett.109.050506} {\bibfield  {journal} {\bibinfo  {journal}
  {Phys. Rev. Lett.}\ }\textbf {\bibinfo {volume} {109}},\ \bibinfo {pages}
  {050506} (\bibinfo {year} {2012})}\BibitemShut {NoStop}%
\bibitem [{\citenamefont {Rist\`e}\ \emph
  {et~al.}(2012{\natexlab{a}})\citenamefont {Rist\`e}, \citenamefont {van
  Leeuwen}, \citenamefont {Ku}, \citenamefont {Lehnert},\ and\ \citenamefont
  {DiCarlo}}]{Riste12}%
  \BibitemOpen
  \bibfield  {author} {\bibinfo {author} {\bibfnamefont {D.}~\bibnamefont
  {Rist\`e}}, \bibinfo {author} {\bibfnamefont {J.~G.}\ \bibnamefont {van
  Leeuwen}}, \bibinfo {author} {\bibfnamefont {H.-S.}\ \bibnamefont {Ku}},
  \bibinfo {author} {\bibfnamefont {K.~W.}\ \bibnamefont {Lehnert}}, \ and\
  \bibinfo {author} {\bibfnamefont {L.}~\bibnamefont {DiCarlo}},\ }\href
  {\doibase 10.1103/PhysRevLett.109.050507} {\bibfield  {journal} {\bibinfo
  {journal} {Phys. Rev. Lett.}\ }\textbf {\bibinfo {volume} {109}},\ \bibinfo
  {pages} {050507} (\bibinfo {year} {2012}{\natexlab{a}})}\BibitemShut
  {NoStop}%
\bibitem [{\citenamefont {Saira}\ \emph {et~al.}(2014)\citenamefont {Saira},
  \citenamefont {Groen}, \citenamefont {Cramer}, \citenamefont {Meretska},
  \citenamefont {de~Lange},\ and\ \citenamefont {DiCarlo}}]{Saira14}%
  \BibitemOpen
  \bibfield  {author} {\bibinfo {author} {\bibfnamefont {O.-P.}\ \bibnamefont
  {Saira}}, \bibinfo {author} {\bibfnamefont {J.~P.}\ \bibnamefont {Groen}},
  \bibinfo {author} {\bibfnamefont {J.}~\bibnamefont {Cramer}}, \bibinfo
  {author} {\bibfnamefont {M.}~\bibnamefont {Meretska}}, \bibinfo {author}
  {\bibfnamefont {G.}~\bibnamefont {de~Lange}}, \ and\ \bibinfo {author}
  {\bibfnamefont {L.}~\bibnamefont {DiCarlo}},\ }\href@noop {} {\bibfield
  {journal} {\bibinfo  {journal} {Phys. Rev. Lett.}\ }\textbf {\bibinfo
  {volume} {112}},\ \bibinfo {pages} {070502} (\bibinfo {year}
  {2014})}\BibitemShut {NoStop}%
\bibitem [{\citenamefont {Reed}\ \emph {et~al.}(2010)\citenamefont {Reed},
  \citenamefont {Johnson}, \citenamefont {Houck}, \citenamefont {DiCarlo},
  \citenamefont {Chow}, \citenamefont {Schuster}, \citenamefont {Frunzio},\
  and\ \citenamefont {Schoelkopf}}]{Reed10b}%
  \BibitemOpen
  \bibfield  {author} {\bibinfo {author} {\bibfnamefont {M.~D.}\ \bibnamefont
  {Reed}}, \bibinfo {author} {\bibfnamefont {B.~R.}\ \bibnamefont {Johnson}},
  \bibinfo {author} {\bibfnamefont {A.~A.}\ \bibnamefont {Houck}}, \bibinfo
  {author} {\bibfnamefont {L.}~\bibnamefont {DiCarlo}}, \bibinfo {author}
  {\bibfnamefont {J.~M.}\ \bibnamefont {Chow}}, \bibinfo {author}
  {\bibfnamefont {D.~I.}\ \bibnamefont {Schuster}}, \bibinfo {author}
  {\bibfnamefont {L.}~\bibnamefont {Frunzio}}, \ and\ \bibinfo {author}
  {\bibfnamefont {R.~J.}\ \bibnamefont {Schoelkopf}},\ }\href@noop {}
  {\bibfield  {journal} {\bibinfo  {journal} {Appl. Phys. Lett.}\ }\textbf
  {\bibinfo {volume} {96}},\ \bibinfo {pages} {203110} (\bibinfo {year}
  {2010})}\BibitemShut {NoStop}%
\bibitem [{\citenamefont {Jeffrey}\ \emph
  {et~al.}(2014{\natexlab{a}})\citenamefont {Jeffrey}, \citenamefont {Sank},
  \citenamefont {Mutus}, \citenamefont {White}, \citenamefont {Kelly},
  \citenamefont {Barends}, \citenamefont {Chen}, \citenamefont {Chen},
  \citenamefont {Chiaro}, \citenamefont {Dunsworth}, \citenamefont {Megrant},
  \citenamefont {O'Malley}, \citenamefont {Neill}, \citenamefont {Roushan},
  \citenamefont {Vainsencher}, \citenamefont {Wenner}, \citenamefont
  {Cleland},\ and\ \citenamefont {Martinis}}]{Jeffrey14}%
  \BibitemOpen
  \bibfield  {author} {\bibinfo {author} {\bibfnamefont {E.}~\bibnamefont
  {Jeffrey}}, \bibinfo {author} {\bibfnamefont {D.}~\bibnamefont {Sank}},
  \bibinfo {author} {\bibfnamefont {J.~Y.}\ \bibnamefont {Mutus}}, \bibinfo
  {author} {\bibfnamefont {T.~C.}\ \bibnamefont {White}}, \bibinfo {author}
  {\bibfnamefont {J.}~\bibnamefont {Kelly}}, \bibinfo {author} {\bibfnamefont
  {R.}~\bibnamefont {Barends}}, \bibinfo {author} {\bibfnamefont
  {Y.}~\bibnamefont {Chen}}, \bibinfo {author} {\bibfnamefont {Z.}~\bibnamefont
  {Chen}}, \bibinfo {author} {\bibfnamefont {B.}~\bibnamefont {Chiaro}},
  \bibinfo {author} {\bibfnamefont {A.}~\bibnamefont {Dunsworth}}, \bibinfo
  {author} {\bibfnamefont {A.}~\bibnamefont {Megrant}}, \bibinfo {author}
  {\bibfnamefont {P.~J.~J.}\ \bibnamefont {O'Malley}}, \bibinfo {author}
  {\bibfnamefont {C.}~\bibnamefont {Neill}}, \bibinfo {author} {\bibfnamefont
  {P.}~\bibnamefont {Roushan}}, \bibinfo {author} {\bibfnamefont
  {A.}~\bibnamefont {Vainsencher}}, \bibinfo {author} {\bibfnamefont
  {J.}~\bibnamefont {Wenner}}, \bibinfo {author} {\bibfnamefont {A.~N.}\
  \bibnamefont {Cleland}}, \ and\ \bibinfo {author} {\bibfnamefont {J.~M.}\
  \bibnamefont {Martinis}},\ }\href@noop {} {\bibfield  {journal} {\bibinfo
  {journal} {Phys. Rev. Lett.}\ }\textbf {\bibinfo {volume} {112}},\ \bibinfo
  {pages} {190504} (\bibinfo {year} {2014}{\natexlab{a}})}\BibitemShut
  {NoStop}%
\bibitem [{\citenamefont {Houck}\ \emph {et~al.}(2008)\citenamefont {Houck},
  \citenamefont {Schreier}, \citenamefont {Johnson}, \citenamefont {Chow},
  \citenamefont {Koch}, \citenamefont {Gambetta}, \citenamefont {Schuster},
  \citenamefont {Frunzio}, \citenamefont {Devoret}, \citenamefont {Girvin},\
  and\ \citenamefont {Schoelkopf}}]{Houck08}%
  \BibitemOpen
  \bibfield  {author} {\bibinfo {author} {\bibfnamefont {A.~A.}\ \bibnamefont
  {Houck}}, \bibinfo {author} {\bibfnamefont {J.~A.}\ \bibnamefont {Schreier}},
  \bibinfo {author} {\bibfnamefont {B.~R.}\ \bibnamefont {Johnson}}, \bibinfo
  {author} {\bibfnamefont {J.~M.}\ \bibnamefont {Chow}}, \bibinfo {author}
  {\bibfnamefont {J.}~\bibnamefont {Koch}}, \bibinfo {author} {\bibfnamefont
  {J.~M.}\ \bibnamefont {Gambetta}}, \bibinfo {author} {\bibfnamefont {D.~I.}\
  \bibnamefont {Schuster}}, \bibinfo {author} {\bibfnamefont {L.}~\bibnamefont
  {Frunzio}}, \bibinfo {author} {\bibfnamefont {M.~H.}\ \bibnamefont
  {Devoret}}, \bibinfo {author} {\bibfnamefont {S.~M.}\ \bibnamefont {Girvin}},
  \ and\ \bibinfo {author} {\bibfnamefont {R.~J.}\ \bibnamefont {Schoelkopf}},\
  }\href@noop {} {\bibfield  {journal} {\bibinfo  {journal} {Phys. Rev. Lett.}\
  }\textbf {\bibinfo {volume} {101}},\ \bibinfo {eid} {080502} (\bibinfo {year}
  {2008})}\BibitemShut {NoStop}%
\bibitem [{\citenamefont {Sears}\ \emph {et~al.}(2012)\citenamefont {Sears},
  \citenamefont {Petrenko}, \citenamefont {Catelani}, \citenamefont {Sun},
  \citenamefont {Paik}, \citenamefont {Kirchmair}, \citenamefont {Frunzio},
  \citenamefont {Glazman}, \citenamefont {Girvin},\ and\ \citenamefont
  {Schoelkopf}}]{Sears12}%
  \BibitemOpen
  \bibfield  {author} {\bibinfo {author} {\bibfnamefont {A.~P.}\ \bibnamefont
  {Sears}}, \bibinfo {author} {\bibfnamefont {A.}~\bibnamefont {Petrenko}},
  \bibinfo {author} {\bibfnamefont {G.}~\bibnamefont {Catelani}}, \bibinfo
  {author} {\bibfnamefont {L.}~\bibnamefont {Sun}}, \bibinfo {author}
  {\bibfnamefont {H.}~\bibnamefont {Paik}}, \bibinfo {author} {\bibfnamefont
  {G.}~\bibnamefont {Kirchmair}}, \bibinfo {author} {\bibfnamefont
  {L.}~\bibnamefont {Frunzio}}, \bibinfo {author} {\bibfnamefont {L.~I.}\
  \bibnamefont {Glazman}}, \bibinfo {author} {\bibfnamefont {S.~M.}\
  \bibnamefont {Girvin}}, \ and\ \bibinfo {author} {\bibfnamefont {R.~J.}\
  \bibnamefont {Schoelkopf}},\ }\href {\doibase 10.1103/PhysRevB.86.180504}
  {\bibfield  {journal} {\bibinfo  {journal} {Phys. Rev. B}\ }\textbf {\bibinfo
  {volume} {86}},\ \bibinfo {pages} {180504} (\bibinfo {year}
  {2012})}\BibitemShut {NoStop}%
\bibitem [{\citenamefont {Jin}\ \emph {et~al.}(2015)\citenamefont {Jin},
  \citenamefont {Kamal}, \citenamefont {Sears}, \citenamefont {Gudmundsen},
  \citenamefont {Hover}, \citenamefont {Miloshi}, \citenamefont {Slattery},
  \citenamefont {Yan}, \citenamefont {Yoder}, \citenamefont {Orlando},
  \citenamefont {Gustavsson},\ and\ \citenamefont {Oliver}}]{Jin15}%
  \BibitemOpen
  \bibfield  {author} {\bibinfo {author} {\bibfnamefont {X.~Y.}\ \bibnamefont
  {Jin}}, \bibinfo {author} {\bibfnamefont {A.}~\bibnamefont {Kamal}}, \bibinfo
  {author} {\bibfnamefont {A.~P.}\ \bibnamefont {Sears}}, \bibinfo {author}
  {\bibfnamefont {T.}~\bibnamefont {Gudmundsen}}, \bibinfo {author}
  {\bibfnamefont {D.}~\bibnamefont {Hover}}, \bibinfo {author} {\bibfnamefont
  {J.}~\bibnamefont {Miloshi}}, \bibinfo {author} {\bibfnamefont
  {R.}~\bibnamefont {Slattery}}, \bibinfo {author} {\bibfnamefont
  {F.}~\bibnamefont {Yan}}, \bibinfo {author} {\bibfnamefont {J.}~\bibnamefont
  {Yoder}}, \bibinfo {author} {\bibfnamefont {T.~P.}\ \bibnamefont {Orlando}},
  \bibinfo {author} {\bibfnamefont {S.}~\bibnamefont {Gustavsson}}, \ and\
  \bibinfo {author} {\bibfnamefont {W.~D.}\ \bibnamefont {Oliver}},\
  }\href@noop {} {\bibfield  {journal} {\bibinfo  {journal} {Phys. Rev. Lett.}\
  }\textbf {\bibinfo {volume} {114}},\ \bibinfo {pages} {240501} (\bibinfo
  {year} {2015})}\BibitemShut {NoStop}%
\bibitem [{\citenamefont {McClure}\ \emph {et~al.}(2016)\citenamefont
  {McClure}, \citenamefont {Paik}, \citenamefont {Bishop}, \citenamefont
  {Steffen}, \citenamefont {Chow},\ and\ \citenamefont {Gambetta}}]{McClure16}%
  \BibitemOpen
  \bibfield  {author} {\bibinfo {author} {\bibfnamefont {D.~T.}\ \bibnamefont
  {McClure}}, \bibinfo {author} {\bibfnamefont {H.}~\bibnamefont {Paik}},
  \bibinfo {author} {\bibfnamefont {L.~S.}\ \bibnamefont {Bishop}}, \bibinfo
  {author} {\bibfnamefont {M.}~\bibnamefont {Steffen}}, \bibinfo {author}
  {\bibfnamefont {J.~M.}\ \bibnamefont {Chow}}, \ and\ \bibinfo {author}
  {\bibfnamefont {J.~M.}\ \bibnamefont {Gambetta}},\ }\href@noop {} {\bibfield
  {journal} {\bibinfo  {journal} {Phys. Rev. Applied}\ }\textbf {\bibinfo
  {volume} {5}},\ \bibinfo {pages} {011001} (\bibinfo {year}
  {2016})}\BibitemShut {NoStop}%
\bibitem [{\citenamefont {Powell}(1964)}]{Powell64}%
  \BibitemOpen
  \bibfield  {author} {\bibinfo {author} {\bibfnamefont {M.~J.~D.}\
  \bibnamefont {Powell}},\ }\href {\doibase 10.1093/comjnl/7.2.155} {\bibfield
  {journal} {\bibinfo  {journal} {The Computer Journal}\ }\textbf {\bibinfo
  {volume} {7}},\ \bibinfo {pages} {155} (\bibinfo {year} {1964})}\BibitemShut
  {NoStop}%
\bibitem [{\citenamefont {Chow}\ \emph {et~al.}(2014)\citenamefont {Chow},
  \citenamefont {Gambetta}, \citenamefont {Magesan}, \citenamefont {Abraham},
  \citenamefont {Cross}, \citenamefont {Johnson}, \citenamefont {Masluk},
  \citenamefont {Ryan}, \citenamefont {Smolin}, \citenamefont {Srinivasan},\
  and\ \citenamefont {Steffen}}]{Chow14}%
  \BibitemOpen
  \bibfield  {author} {\bibinfo {author} {\bibfnamefont {J.~M.}\ \bibnamefont
  {Chow}}, \bibinfo {author} {\bibfnamefont {J.~M.}\ \bibnamefont {Gambetta}},
  \bibinfo {author} {\bibfnamefont {E.}~\bibnamefont {Magesan}}, \bibinfo
  {author} {\bibfnamefont {D.~W.}\ \bibnamefont {Abraham}}, \bibinfo {author}
  {\bibfnamefont {A.~W.}\ \bibnamefont {Cross}}, \bibinfo {author}
  {\bibfnamefont {B.~R.}\ \bibnamefont {Johnson}}, \bibinfo {author}
  {\bibfnamefont {N.~A.}\ \bibnamefont {Masluk}}, \bibinfo {author}
  {\bibfnamefont {C.~A.}\ \bibnamefont {Ryan}}, \bibinfo {author}
  {\bibfnamefont {J.~A.}\ \bibnamefont {Smolin}}, \bibinfo {author}
  {\bibfnamefont {S.~J.}\ \bibnamefont {Srinivasan}}, \ and\ \bibinfo {author}
  {\bibfnamefont {M.}~\bibnamefont {Steffen}},\ }\href@noop {} {\bibfield
  {journal} {\bibinfo  {journal} {Nat.\ Commun.}\ }\textbf {\bibinfo {volume}
  {5}},\ \bibinfo {pages} {4015} (\bibinfo {year} {2014})}\BibitemShut
  {NoStop}%
\bibitem [{\citenamefont {O'Brien}()}]{oBrien16}%
  \BibitemOpen
  \bibfield  {author} {\bibinfo {author} {\bibfnamefont {T.~E.}\ \bibnamefont
  {O'Brien}},\ }\href@noop {} {}\bibinfo {howpublished} {in preparation
  (2016)}\BibitemShut {NoStop}%
\bibitem [{\citenamefont {Tomita}\ and\ \citenamefont
  {Svore}(2014)}]{Tomita14}%
  \BibitemOpen
  \bibfield  {author} {\bibinfo {author} {\bibfnamefont {Y.}~\bibnamefont
  {Tomita}}\ and\ \bibinfo {author} {\bibfnamefont {K.~M.}\ \bibnamefont
  {Svore}},\ }\href@noop {} {\bibfield  {journal} {\bibinfo  {journal} {Phys.
  Rev. A}\ }\textbf {\bibinfo {volume} {90}},\ \bibinfo {pages} {062320}
  (\bibinfo {year} {2014})}\BibitemShut {NoStop}%
\bibitem [{\citenamefont {Schuster}\ \emph {et~al.}(2007)\citenamefont
  {Schuster}, \citenamefont {Houck}, \citenamefont {Schreier}, \citenamefont
  {Wallraff}, \citenamefont {Gambetta}, \citenamefont {Blais}, \citenamefont
  {Frunzio}, \citenamefont {Majer}, \citenamefont {Devoret}, \citenamefont
  {Givin},\ and\ \citenamefont {Schoelkopf}}]{Schuster07}%
  \BibitemOpen
  \bibfield  {author} {\bibinfo {author} {\bibfnamefont {D.~I.}\ \bibnamefont
  {Schuster}}, \bibinfo {author} {\bibfnamefont {A.~A.}\ \bibnamefont {Houck}},
  \bibinfo {author} {\bibfnamefont {J.~A.}\ \bibnamefont {Schreier}}, \bibinfo
  {author} {\bibfnamefont {A.}~\bibnamefont {Wallraff}}, \bibinfo {author}
  {\bibfnamefont {J.~M.}\ \bibnamefont {Gambetta}}, \bibinfo {author}
  {\bibfnamefont {A.}~\bibnamefont {Blais}}, \bibinfo {author} {\bibfnamefont
  {L.}~\bibnamefont {Frunzio}}, \bibinfo {author} {\bibfnamefont
  {J.}~\bibnamefont {Majer}}, \bibinfo {author} {\bibfnamefont {M.~H.}\
  \bibnamefont {Devoret}}, \bibinfo {author} {\bibfnamefont {S.~M.}\
  \bibnamefont {Givin}}, \ and\ \bibinfo {author} {\bibfnamefont {R.~J.}\
  \bibnamefont {Schoelkopf}},\ }\href@noop {} {\bibfield  {journal} {\bibinfo
  {journal} {Nature}\ }\textbf {\bibinfo {volume} {445}},\ \bibinfo {pages}
  {515} (\bibinfo {year} {2007})}\BibitemShut {NoStop}%
\bibitem [{SOM()}]{SOMdepletion}%
  \BibitemOpen
  \href@noop {} {}\bibinfo {howpublished} {See supplemental material.}\BibitemShut {Stop}%
\bibitem [{\citenamefont {Ryan}\ \emph {et~al.}(2015)\citenamefont {Ryan},
  \citenamefont {Johnson}, \citenamefont {Gambetta}, \citenamefont {Chow},
  \citenamefont {da~Silva}, \citenamefont {Dial},\ and\ \citenamefont
  {Ohki}}]{Ryan15}%
  \BibitemOpen
  \bibfield  {author} {\bibinfo {author} {\bibfnamefont {C.~A.}\ \bibnamefont
  {Ryan}}, \bibinfo {author} {\bibfnamefont {B.~R.}\ \bibnamefont {Johnson}},
  \bibinfo {author} {\bibfnamefont {J.~M.}\ \bibnamefont {Gambetta}}, \bibinfo
  {author} {\bibfnamefont {J.~M.}\ \bibnamefont {Chow}}, \bibinfo {author}
  {\bibfnamefont {M.~P.}\ \bibnamefont {da~Silva}}, \bibinfo {author}
  {\bibfnamefont {O.~E.}\ \bibnamefont {Dial}}, \ and\ \bibinfo {author}
  {\bibfnamefont {T.~A.}\ \bibnamefont {Ohki}},\ }\href@noop {} {\bibfield
  {journal} {\bibinfo  {journal} {Phys. Rev. A}\ }\textbf {\bibinfo {volume}
  {91}},\ \bibinfo {pages} {022118} (\bibinfo {year} {2015})}\BibitemShut
  {NoStop}%
\bibitem [{\citenamefont {Magesan}\ \emph {et~al.}(2015)\citenamefont
  {Magesan}, \citenamefont {Gambetta}, \citenamefont {C\'orcoles},\ and\
  \citenamefont {Chow}}]{Magesan15}%
  \BibitemOpen
  \bibfield  {author} {\bibinfo {author} {\bibfnamefont {E.}~\bibnamefont
  {Magesan}}, \bibinfo {author} {\bibfnamefont {J.~M.}\ \bibnamefont
  {Gambetta}}, \bibinfo {author} {\bibfnamefont {A.~D.}\ \bibnamefont
  {C\'orcoles}}, \ and\ \bibinfo {author} {\bibfnamefont {J.~M.}\ \bibnamefont
  {Chow}},\ }\href@noop {} {\bibfield  {journal} {\bibinfo  {journal} {Phys.
  Rev. Lett.}\ }\textbf {\bibinfo {volume} {114}},\ \bibinfo {pages} {200501}
  (\bibinfo {year} {2015})}\BibitemShut {NoStop}%
\bibitem [{\citenamefont {Chow}\ \emph {et~al.}(2010)\citenamefont {Chow},
  \citenamefont {DiCarlo}, \citenamefont {Gambetta}, \citenamefont {Motzoi},
  \citenamefont {Frunzio}, \citenamefont {Girvin},\ and\ \citenamefont
  {Schoelkopf}}]{Chow10b}%
  \BibitemOpen
  \bibfield  {author} {\bibinfo {author} {\bibfnamefont {J.~M.}\ \bibnamefont
  {Chow}}, \bibinfo {author} {\bibfnamefont {L.}~\bibnamefont {DiCarlo}},
  \bibinfo {author} {\bibfnamefont {J.~M.}\ \bibnamefont {Gambetta}}, \bibinfo
  {author} {\bibfnamefont {F.}~\bibnamefont {Motzoi}}, \bibinfo {author}
  {\bibfnamefont {L.}~\bibnamefont {Frunzio}}, \bibinfo {author} {\bibfnamefont
  {S.~M.}\ \bibnamefont {Girvin}}, \ and\ \bibinfo {author} {\bibfnamefont
  {R.~J.}\ \bibnamefont {Schoelkopf}},\ }\href@noop {} {\bibfield  {journal}
  {\bibinfo  {journal} {Phys. Rev. A}\ }\textbf {\bibinfo {volume} {82}},\
  \bibinfo {pages} {040305} (\bibinfo {year} {2010})}\BibitemShut {NoStop}%
\bibitem [{\citenamefont {Reed}(2013)}]{ReedPhD13}%
  \BibitemOpen
  \bibfield  {author} {\bibinfo {author} {\bibfnamefont {M.}~\bibnamefont
  {Reed}},\ }\emph {\bibinfo {title} {Entanglement and quantum error correction
  with superconducting qubits}},\ \href@noop {} {\bibinfo {type} {Ph{D}
  {D}issertation}},\ \bibinfo  {school} {Yale University} (\bibinfo {year}
  {2013})\BibitemShut {NoStop}%
\bibitem [{\citenamefont {Jeffrey}\ \emph
  {et~al.}(2014{\natexlab{b}})\citenamefont {Jeffrey}, \citenamefont {Sank},
  \citenamefont {Mutus}, \citenamefont {White}, \citenamefont {Kelly},
  \citenamefont {Barends}, \citenamefont {Chen}, \citenamefont {Chen},
  \citenamefont {Chiaro}, \citenamefont {Dunsworth}, \citenamefont {Megrant},
  \citenamefont {O'Malley}, \citenamefont {Neill}, \citenamefont {Roushan},
  \citenamefont {Vainsencher}, \citenamefont {Wenner}, \citenamefont
  {Cleland},\ and\ \citenamefont {Martinis}}]{Sank14}%
  \BibitemOpen
  \bibfield  {author} {\bibinfo {author} {\bibfnamefont {E.}~\bibnamefont
  {Jeffrey}}, \bibinfo {author} {\bibfnamefont {D.}~\bibnamefont {Sank}},
  \bibinfo {author} {\bibfnamefont {J.~Y.}\ \bibnamefont {Mutus}}, \bibinfo
  {author} {\bibfnamefont {T.~C.}\ \bibnamefont {White}}, \bibinfo {author}
  {\bibfnamefont {J.}~\bibnamefont {Kelly}}, \bibinfo {author} {\bibfnamefont
  {R.}~\bibnamefont {Barends}}, \bibinfo {author} {\bibfnamefont
  {Y.}~\bibnamefont {Chen}}, \bibinfo {author} {\bibfnamefont {Z.}~\bibnamefont
  {Chen}}, \bibinfo {author} {\bibfnamefont {B.}~\bibnamefont {Chiaro}},
  \bibinfo {author} {\bibfnamefont {A.}~\bibnamefont {Dunsworth}}, \bibinfo
  {author} {\bibfnamefont {A.}~\bibnamefont {Megrant}}, \bibinfo {author}
  {\bibfnamefont {P.~J.~J.}\ \bibnamefont {O'Malley}}, \bibinfo {author}
  {\bibfnamefont {C.}~\bibnamefont {Neill}}, \bibinfo {author} {\bibfnamefont
  {P.}~\bibnamefont {Roushan}}, \bibinfo {author} {\bibfnamefont
  {A.}~\bibnamefont {Vainsencher}}, \bibinfo {author} {\bibfnamefont
  {J.}~\bibnamefont {Wenner}}, \bibinfo {author} {\bibfnamefont {A.~N.}\
  \bibnamefont {Cleland}}, \ and\ \bibinfo {author} {\bibfnamefont {J.~M.}\
  \bibnamefont {Martinis}},\ }\href@noop {} {\bibfield  {journal} {\bibinfo
  {journal} {Phys. Rev. Lett.}\ }\textbf {\bibinfo {volume} {112}} (\bibinfo
  {year} {2014}{\natexlab{b}})}\BibitemShut {NoStop}%
\bibitem [{\citenamefont {Fowler}\ \emph {et~al.}(2014)\citenamefont {Fowler},
  \citenamefont {Sank}, \citenamefont {Kelly}, \citenamefont {Barends},\ and\
  \citenamefont {Martinis}}]{Fowler14}%
  \BibitemOpen
  \bibfield  {author} {\bibinfo {author} {\bibfnamefont {A.~G.}\ \bibnamefont
  {Fowler}}, \bibinfo {author} {\bibfnamefont {D.}~\bibnamefont {Sank}},
  \bibinfo {author} {\bibfnamefont {J.}~\bibnamefont {Kelly}}, \bibinfo
  {author} {\bibfnamefont {R.}~\bibnamefont {Barends}}, \ and\ \bibinfo
  {author} {\bibfnamefont {J.~M.}\ \bibnamefont {Martinis}},\ }\href@noop {}
  {\bibfield  {journal} {\bibinfo  {journal} {arXiv:1405.1454}\ } (\bibinfo
  {year} {2014})}\BibitemShut {NoStop}%
\bibitem [{\citenamefont {Frisk~Kockum}\ \emph {et~al.}(2012)\citenamefont
  {Frisk~Kockum}, \citenamefont {Tornberg},\ and\ \citenamefont
  {Johansson}}]{FriskKockum12}%
  \BibitemOpen
  \bibfield  {author} {\bibinfo {author} {\bibfnamefont {A.}~\bibnamefont
  {Frisk~Kockum}}, \bibinfo {author} {\bibfnamefont {L.}~\bibnamefont
  {Tornberg}}, \ and\ \bibinfo {author} {\bibfnamefont {G.}~\bibnamefont
  {Johansson}},\ }\href@noop {} {\bibfield  {journal} {\bibinfo  {journal}
  {Phys. Rev. A}\ }\textbf {\bibinfo {volume} {85}},\ \bibinfo {pages} {052318}
  (\bibinfo {year} {2012})}\BibitemShut {NoStop}%
\bibitem [{\citenamefont {Rist\`e}\ \emph
  {et~al.}(2012{\natexlab{b}})\citenamefont {Rist\`e}, \citenamefont {Bultink},
  \citenamefont {Lehnert},\ and\ \citenamefont {DiCarlo}}]{Riste12b}%
  \BibitemOpen
  \bibfield  {author} {\bibinfo {author} {\bibfnamefont {D.}~\bibnamefont
  {Rist\`e}}, \bibinfo {author} {\bibfnamefont {C.~C.}\ \bibnamefont
  {Bultink}}, \bibinfo {author} {\bibfnamefont {K.~W.}\ \bibnamefont
  {Lehnert}}, \ and\ \bibinfo {author} {\bibfnamefont {L.}~\bibnamefont
  {DiCarlo}},\ }\href@noop {} {\bibfield  {journal} {\bibinfo  {journal} {Phys.
  Rev. Lett.}\ }\textbf {\bibinfo {volume} {109}},\ \bibinfo {pages} {240502}
  (\bibinfo {year} {2012}{\natexlab{b}})}\BibitemShut {NoStop}%
\bibitem [{\citenamefont {Rol}()}]{Rol16}%
  \BibitemOpen
  \bibfield  {author} {\bibinfo {author} {\bibfnamefont {M.~A.}\ \bibnamefont
  {Rol}},\ }\href@noop {} {}\bibinfo {howpublished} {in preparation
  (2016)}\BibitemShut {NoStop}%
\bibitem [{\citenamefont {Kelly}\ \emph {et~al.}(2016)\citenamefont {Kelly},
  \citenamefont {Barends}, \citenamefont {Fowler}, \citenamefont {Megrant},
  \citenamefont {Jeffrey}, \citenamefont {White}, \citenamefont {Sank},
  \citenamefont {Mutus}, \citenamefont {Campbell}, \citenamefont {Chen},
  \citenamefont {Chen}, \citenamefont {Chiaro}, \citenamefont {Dunsworth},
  \citenamefont {Lucero}, \citenamefont {Neeley}, \citenamefont {Neill},
  \citenamefont {O'Malley}, \citenamefont {Quintana}, \citenamefont {Roushan},
  \citenamefont {Vainsencher}, \citenamefont {Wenner},\ and\ \citenamefont
  {Martinis}}]{Kelly16}%
  \BibitemOpen
  \bibfield  {author} {\bibinfo {author} {\bibfnamefont {J.}~\bibnamefont
  {Kelly}}, \bibinfo {author} {\bibfnamefont {R.}~\bibnamefont {Barends}},
  \bibinfo {author} {\bibfnamefont {A.~G.}\ \bibnamefont {Fowler}}, \bibinfo
  {author} {\bibfnamefont {A.}~\bibnamefont {Megrant}}, \bibinfo {author}
  {\bibfnamefont {E.}~\bibnamefont {Jeffrey}}, \bibinfo {author} {\bibfnamefont
  {T.~C.}\ \bibnamefont {White}}, \bibinfo {author} {\bibfnamefont
  {D.}~\bibnamefont {Sank}}, \bibinfo {author} {\bibfnamefont {J.~Y.}\
  \bibnamefont {Mutus}}, \bibinfo {author} {\bibfnamefont {B.}~\bibnamefont
  {Campbell}}, \bibinfo {author} {\bibfnamefont {Y.}~\bibnamefont {Chen}},
  \bibinfo {author} {\bibfnamefont {Z.}~\bibnamefont {Chen}}, \bibinfo {author}
  {\bibfnamefont {B.}~\bibnamefont {Chiaro}}, \bibinfo {author} {\bibfnamefont
  {A.}~\bibnamefont {Dunsworth}}, \bibinfo {author} {\bibfnamefont
  {E.}~\bibnamefont {Lucero}}, \bibinfo {author} {\bibfnamefont
  {M.}~\bibnamefont {Neeley}}, \bibinfo {author} {\bibfnamefont
  {C.}~\bibnamefont {Neill}}, \bibinfo {author} {\bibfnamefont {P.~J.~J.}\
  \bibnamefont {O'Malley}}, \bibinfo {author} {\bibfnamefont {C.}~\bibnamefont
  {Quintana}}, \bibinfo {author} {\bibfnamefont {P.}~\bibnamefont {Roushan}},
  \bibinfo {author} {\bibfnamefont {A.}~\bibnamefont {Vainsencher}}, \bibinfo
  {author} {\bibfnamefont {J.}~\bibnamefont {Wenner}}, \ and\ \bibinfo {author}
  {\bibfnamefont {J.~M.}\ \bibnamefont {Martinis}},\ }\href@noop {} {\bibfield
  {journal} {\bibinfo  {journal} {arXiv:1603.03082}\ } (\bibinfo {year}
  {2016})}\BibitemShut {NoStop}%
\end{thebibliography}

\begin{thebibliography}{4}%
\makeatletter
\providecommand \@ifxundefined [1]{%
 \@ifx{#1\undefined}
}%
\providecommand \@ifnum [1]{%
 \ifnum #1\expandafter \@firstoftwo
 \else \expandafter \@secondoftwo
 \fi
}%
\providecommand \@ifx [1]{%
 \ifx #1\expandafter \@firstoftwo
 \else \expandafter \@secondoftwo
 \fi
}%
\providecommand \natexlab [1]{#1}%
\providecommand \enquote  [1]{``#1''}%
\providecommand \bibnamefont  [1]{#1}%
\providecommand \bibfnamefont [1]{#1}%
\providecommand \citenamefont [1]{#1}%
\providecommand \href@noop [0]{\@secondoftwo}%
\providecommand \href [0]{\begingroup \@sanitize@url \@href}%
\providecommand \@href[1]{\@@startlink{#1}\@@href}%
\providecommand \@@href[1]{\endgroup#1\@@endlink}%
\providecommand \@sanitize@url [0]{\catcode `\\12\catcode `\$12\catcode
  `\&12\catcode `\#12\catcode `\^12\catcode `\_12\catcode `\%12\relax}%
\providecommand \@@startlink[1]{}%
\providecommand \@@endlink[0]{}%
\providecommand \url  [0]{\begingroup\@sanitize@url \@url }%
\providecommand \@url [1]{\endgroup\@href {#1}{\urlprefix }}%
\providecommand \urlprefix  [0]{URL }%
\providecommand \Eprint [0]{\href }%
\providecommand \doibase [0]{http://dx.doi.org/}%
\providecommand \selectlanguage [0]{\@gobble}%
\providecommand \bibinfo  [0]{\@secondoftwo}%
\providecommand \bibfield  [0]{\@secondoftwo}%
\providecommand \translation [1]{[#1]}%
\providecommand \BibitemOpen [0]{}%
\providecommand \bibitemStop [0]{}%
\providecommand \bibitemNoStop [0]{.\EOS\space}%
\providecommand \EOS [0]{\spacefactor3000\relax}%
\providecommand \BibitemShut  [1]{\csname bibitem#1\endcsname}%
\let\auto@bib@innerbib\@empty
\bibitem [{\citenamefont {Rist\`{e}}\ \emph {et~al.}(2015)\citenamefont
  {Rist\`{e}}, \citenamefont {Poletto}, \citenamefont {Huang}, \citenamefont
  {Bruno}, \citenamefont {Vesterinen}, \citenamefont {Saira},\ and\
  \citenamefont {DiCarlo}}]{Riste15_b}%
  \BibitemOpen
  \bibfield  {author} {\bibinfo {author} {\bibfnamefont {D.}~\bibnamefont
  {Rist\`{e}}}, \bibinfo {author} {\bibfnamefont {S.}~\bibnamefont {Poletto}},
  \bibinfo {author} {\bibfnamefont {M.~Z.}\ \bibnamefont {Huang}}, \bibinfo
  {author} {\bibfnamefont {A.}~\bibnamefont {Bruno}}, \bibinfo {author}
  {\bibfnamefont {V.}~\bibnamefont {Vesterinen}}, \bibinfo {author}
  {\bibfnamefont {O.~P.}\ \bibnamefont {Saira}}, \ and\ \bibinfo {author}
  {\bibfnamefont {L.}~\bibnamefont {DiCarlo}},\ }\href@noop {} {\bibfield
  {journal} {\bibinfo  {journal} {Nat.\ Commun.}\ }\textbf {\bibinfo {volume}
  {{6}}} (\bibinfo {year} {{2015}})}\BibitemShut {NoStop}%
\bibitem [{\citenamefont {Schuster}\ \emph {et~al.}(2007)\citenamefont
  {Schuster}, \citenamefont {Houck}, \citenamefont {Schreier}, \citenamefont
  {Wallraff}, \citenamefont {Gambetta}, \citenamefont {Blais}, \citenamefont
  {Frunzio}, \citenamefont {Majer}, \citenamefont {Devoret}, \citenamefont
  {Givin},\ and\ \citenamefont {Schoelkopf}}]{Schuster07_b}%
  \BibitemOpen
  \bibfield  {author} {\bibinfo {author} {\bibfnamefont {D.~I.}\ \bibnamefont
  {Schuster}}, \bibinfo {author} {\bibfnamefont {A.~A.}\ \bibnamefont {Houck}},
  \bibinfo {author} {\bibfnamefont {J.~A.}\ \bibnamefont {Schreier}}, \bibinfo
  {author} {\bibfnamefont {A.}~\bibnamefont {Wallraff}}, \bibinfo {author}
  {\bibfnamefont {J.~M.}\ \bibnamefont {Gambetta}}, \bibinfo {author}
  {\bibfnamefont {A.}~\bibnamefont {Blais}}, \bibinfo {author} {\bibfnamefont
  {L.}~\bibnamefont {Frunzio}}, \bibinfo {author} {\bibfnamefont
  {J.}~\bibnamefont {Majer}}, \bibinfo {author} {\bibfnamefont {M.~H.}\
  \bibnamefont {Devoret}}, \bibinfo {author} {\bibfnamefont {S.~M.}\
  \bibnamefont {Givin}}, \ and\ \bibinfo {author} {\bibfnamefont {R.~J.}\
  \bibnamefont {Schoelkopf}},\ }\href@noop {} {\bibfield  {journal} {\bibinfo
  {journal} {Nature}\ }\textbf {\bibinfo {volume} {445}},\ \bibinfo {pages}
  {515} (\bibinfo {year} {2007})}\BibitemShut {NoStop}%
\bibitem [{\citenamefont {Nielsen}\ and\ \citenamefont
  {Chuang}(2000)}]{Nielsen00}%
  \BibitemOpen
  \bibfield  {author} {\bibinfo {author} {\bibfnamefont {M.~A.}\ \bibnamefont
  {Nielsen}}\ and\ \bibinfo {author} {\bibfnamefont {I.~L.}\ \bibnamefont
  {Chuang}},\ }\href@noop {} {\emph {\bibinfo {title} {Quantum Computation and
  Quantum Information}}}\ (\bibinfo  {publisher} {Cambridge University Press},\
  \bibinfo {address} {Cambridge},\ \bibinfo {year} {2000})\BibitemShut
  {NoStop}%
\bibitem [{\citenamefont {Frisk~Kockum}\ \emph {et~al.}(2012)\citenamefont
  {Frisk~Kockum}, \citenamefont {Tornberg},\ and\ \citenamefont
  {Johansson}}]{FriskKockum12_b}%
  \BibitemOpen
  \bibfield  {author} {\bibinfo {author} {\bibfnamefont {A.}~\bibnamefont
  {Frisk~Kockum}}, \bibinfo {author} {\bibfnamefont {L.}~\bibnamefont
  {Tornberg}}, \ and\ \bibinfo {author} {\bibfnamefont {G.}~\bibnamefont
  {Johansson}},\ }\href@noop {} {\bibfield  {journal} {\bibinfo  {journal}
  {Phys. Rev. A}\ }\textbf {\bibinfo {volume} {85}},\ \bibinfo {pages} {052318}
  (\bibinfo {year} {2012})}\BibitemShut {NoStop}%
\end{thebibliography}
%

\end{document}